\newif\ifconfver
\newif\ifcutshort      %this level shortens the equations
\cutshorttrue

\newif\ifcutshortlvltwo  %this level takes out some examples, figs., and sim.
\cutshortlvltwofalse

\ifconfver
    \documentclass[10pt,twocolumn,twoside]{IEEEtran}
\else
    \documentclass[11pt,draftcls,onecolumn]{IEEEtran}
\fi

\usepackage{array}
\usepackage{cite}
\usepackage{calc}
\usepackage{caption}
\usepackage{graphicx}
\usepackage{psfrag}
\usepackage[tight,footnotesize]{subfigure}
\usepackage{stfloats}
\usepackage{url}
\usepackage{subfig}
\usepackage{longtable}
\usepackage{amsmath,amsfonts,amssymb,amsbsy,bm,paralist,theorem,ifthen,color}
\usepackage{algorithmic,algorithm}
\usepackage[dvips]{epsfig}
\usepackage[dvips]{graphicx}

\newcommand\Wc{\ensuremath{\mathcal{W}}}

\newcommand\Vc{\ensuremath{{\mathcal{V}}}}

\newcommand\Lc{\ensuremath{{\mathcal{L}}}}
\newcommand\Pc{\ensuremath{{\mathcal{P}}}}
\newcommand\Ac{\ensuremath{{\mathcal{A}}}}
\newcommand\Nc{\ensuremath{{\mathcal{N}}}}
\newcommand\xb{\ensuremath{{\bm x}}}

\newcommand\wb{\ensuremath{{\bm w}}}
\newcommand\yb{\ensuremath{{\bm y}}}

\newcommand\Ab{\ensuremath{{\bm A}}}
\newcommand\ab{\ensuremath{{\bm a}}}

\newcommand\bb{\ensuremath{{\bm b}}}

\newcommand\cb{\ensuremath{{\bf c}}}
\newcommand\db{\ensuremath{{\bm d}}}

\newcommand\lambdab{\ensuremath{{\bm \lambda}}}

\newcommand\zerob{\ensuremath{{\bm 0}}}

\newcommand\oneb{\ensuremath{{\bf 1}}}

\newcommand\Xc{\ensuremath{{\mathcal{X}}}}

\newtheorem{Lemma}{Lemma}

\newtheorem{Theorem}{Theorem}

\newtheorem{Assumption}{Assumption}

\ifconfver

\else

\fi

\begin{document}

\bibliographystyle{IEEEtran}

%\title{A Dual Consensus ADMM Method for Distributed Optimization : Iteration Complexity and Application to Smart Grid Load Control}
\title{Asynchronous ADMM for Distributed Optimization with Heterogeneous Processors}
\title{Asynchronous Distributed ADMM for Large-Scale Optimization- Part II: Linear Convergence Analysis and Numerical Performance}
%\title{Distributed Delayed Alternating Direction Method of Multipliers for Big Data Optimization- Part II: Convergence Rate Analysis}

\ifconfver \else {\linespread{1.1} \rm \fi

\author{\vspace{0.8cm}Tsung-Hui Chang$^\star$, Wei-Cheng Liao$^\S$, Mingyi Hong$^\dag$ and Xiangfeng Wang$^\ddag$ \\
%\thanks{%Part of this work was submitted to NIPS 2015, Montreal, Canada, Dec. 7-12, 2015 \cite{ChangNIPS15}.
%\thanks{%Copyright (c) 2012 IEEE. Personal use of this material is permitted. However, permission to use this material for any other purposes must be obtained
%from the IEEE by sending a request to pubs-permissions@ieee.org.
%The work of Tsung-Hui Chang is supported by Ministry of Science and Technology, Taiwan (R.O.C.), under Grant
%NSC 102-2221-E-011-005-MY3. }
\thanks{$^\star$Tsung-Hui Chang is the corresponding author. Address:
School of Science and Engineering, The Chinese University of Hong Kong, Shenzhen, China 518172,
%Department of Electronic and Computer Engineering, National Taiwan University of Science and Technology, Taipei 10607, Taiwan, (R.O.C.). 
E-mail:
tsunghui.chang@ieee.org. }
\thanks{$^\dag$Wei-Cheng Liao is with Department of Electrical and Computer Engineering, University of Minnesota, Minneapolis, MN 55455, USA, E-mail: mhong@umn.edu}
\thanks{$^\dag$Mingyi Hong is with Department of Industrial and Manufacturing Systems Engineering, Iowa State University, Ames, 50011, USA, E-mail: mingyi@iastate.edu}
\thanks{$^\ddag$Xiangfeng Wang is with Shanghai Key Lab for Trustworthy Computing, Software Engineering Institute, East China Normal University, Shanghai, 200062, China, E-mail: xfwang@sei.ecnu.edu.cn}
}
 \maketitle
\vspace{-0.5cm}
\begin{abstract}
The alternating direction method of multipliers (ADMM) has been recognized as a versatile approach for solving modern large-scale machine learning and signal processing problems efficiently. When the data size and/or the problem dimension is large, a distributed version of ADMM can be used, which is capable of distributing the computation load and the data set to a network of computing nodes. Unfortunately, a direct synchronous implementation of such algorithm does not scale well with the problem size, as the algorithm speed is limited by the slowest computing nodes.
To address this issue, in a companion paper, we have proposed an asynchronous distributed ADMM (AD-ADMM) and studied its worst-case convergence conditions.
In this paper, we further the study by characterizing the conditions under which the AD-ADMM achieves linear convergence.
Our conditions as well as the resulting linear rates reveal the impact that various algorithm parameters, network delay and network size have on the algorithm performance. To demonstrate the superior time efficiency of the proposed AD-ADMM, we test the AD-ADMM on a high-performance computer cluster by solving a large-scale logistic regression problem.
\end{abstract}

%---------------------------------------------------------------------------
\ifconfver \else
\newpage
\fi

\ifconfver \else \IEEEpeerreviewmaketitle} \fi

%\IEEEpeerreviewmaketitle
\vspace{-0.0cm}
\section{Introduction}\label{sec: intro}
\vspace{-0.0cm}
%In this paper, we are interested in solving
Consider the following optimization problem
%\begin{subequations}
\begin{align}\label{eqn: original problem}
  \min_{\substack{\xb\in \mathbb{R}^n}}~ &\sum_{i=1}^N f_i(\xb) + h(\xb),
%  \\
%  \text{s.t.}~ & \xb_i =\xb_0~\forall i=1,\ldots,N,\label{eqn: consensus problem C1}
\end{align}
%\end{subequations}
where each $f_i :\mathbb{R}^{n} \rightarrow \mathbb{R}$ is the cost function and $h:\mathbb{R}^{n} \rightarrow  \mathbb{R}\cup\{\infty\}$ is a non-smooth, convex regularization function.
The regularization function is used for obtaining structured solutions (e.g., sparsity) and/or is an indicator function which enforces $\xb$ to lie in a constraint set \cite[Section 5]{BoydADMM11}.
Many important statistical learning problems can be formulated as problem \eqref{eqn: original problem}, including, for example, the LASSO problem \cite{Tibshirani05_fusedLASSO}, logistic regression (LR) problem \cite{Liu2009}, support vector machine (SVM) \cite{Hastie2001Book} and the sparse principal component analysis (PCA) problem \cite{Richtarikspca12}, to name a few.

Distributed optimization algorithms that can scale well with large-scale instances of \eqref{eqn: original problem}
have drawn significant attention in recent years \cite{BertsekasADMM,BoydADMM11,NiuNIPS11,AgarwalNIPS11,MLiPS14,MLiNIPS14,LiuWright15, meisam14nips, scutari13decomposition}.
Our interest in this paper lies in the distributed optimization method based on the alternating direction method of multipliers (ADMM) \cite[Section 7.1.1]{BoydADMM11}.
The ADMM is a convenient approach of distributing the computation load of a very large-scale problem to a network of computing nodes. Specifically, consider a computer network with a star topology, where one master node coordinates the computation of a set of $N$ distributed workers. Based on a consensus formulation, the distributed ADMM partitions the original problem into $N$ subproblems, each of which contains either a small set of training samples or a subset of the learning parameters. At each iteration, the distributed workers solve the subproblems based on the local data and send the variable information to the master, who summarizes the variable information and broadcasts it back the workers.
Through such iterative variable update and information exchange, the large-scale learning problem can be solved in a distributed and parallel manner.

The convergence conditions of the distributed ADMM have been extensively studied; see \cite{BertsekasADMM,BoydADMM11,He12ADMMite,Hong15noncvxadmm,DengYin2013J,ShiLing2013J,ChangTSP15_CADMM,JakoveticTAC15}.
For example, for general convex problems, references \cite{BertsekasADMM,BoydADMM11} showed that the ADMM is guaranteed to converge to an optimal solution and \cite{He12ADMMite} showed that the ADMM has a worst-case $\mathcal{O}(1/k)$ convergence rate, where $k$ is the iteration number.
Considering non-convex problems with smooth $f_i$'s, reference \cite{Hong15noncvxadmm} presented conditions for which the distributed ADMM converges to the set of Karush-Kuhn-Tucker (KKT) points. For problems with strongly convex and smooth $f_i$'s or problems satisfying certain error bound condition, references \cite{DengYin2013J} and \cite{HongLuo2013} respectively showed that the ADMM can even exhibit a linear convergence rate. References \cite{ShiLing2013J,ChangTSP15_CADMM,JakoveticTAC15} also showed similar linear convergence conditions for some variants of distributed ADMM in a network with a general topology.
However, the distributed ADMM in \cite{BoydADMM11,Hong15noncvxadmm} have assumed a synchronous network, where at each iteration, the master always waits until all the workers report their variable information. Unfortunately,
such synchronous protocol does not scale well with the problem size, as the algorithm speed is determined by the ``slowest" workers. To improve the time efficiency, the works \cite{Zhang14ACADMM,ChangAsyncadmm15_p1} have generalized the distributed ADMM to an asynchronous network. Specifically, in the asynchronous distributed ADMM (AD-ADMM) proposed in \cite{ChangAsyncadmm15_p1,Zhang14ACADMM}, the master does not necessarily wait for all the workers. Instead, the master updates its variable whenever it receives the variable information from a partial set of the workers. This prevents the master and speedy workers from spending most of the time waiting and consequently can improve the time efficiency of distributed optimization.
Theoretically, it has been shown in \cite{ChangAsyncadmm15_p1} that the AD-ADMM is guaranteed to converge (to a KKT point) even for non-convex problem \eqref{eqn: original problem}, under a bounded delay assumption only.
%
%Comparing to the existing stochastic convergence results in \cite{Zhang14ACADMM,ErminWei2013arxiv} which are obtained under some statistical assumption on the workers, the convergence results in \cite{ChangAsyncadmm15_p1} is stronger as it characterizes the worst-case convergence conditions. In addition, the convergence results in \cite{ChangAsyncadmm15_p1} also apply for non-convex problem \eqref{eqn: original problem}, whereas \cite{Zhang14ACADMM,ErminWei2013arxiv} considered convex problems only.

The contributions of this paper are twofold. Firstly, beyond the convergence analysis in \cite{ChangAsyncadmm15_p1}, we further present the conditions for which the AD-ADMM can exhibit a linear convergence rate.
Specifically, we show that for problem \eqref{eqn: original problem} with some structured convex $f_i$'s (e.g., strongly convex), the augmented Lagrangian function of the AD-ADMM can decrease by a constant fraction in every iteration of the algorithm, as long as the algorithm parameters are chosen appropriately according to the network delay. We give explicit expressions on the linear convergence conditions and the linear rate, which illustrate how the algorithm and network parameters impact on the algorithm performance.
%Since, as shown in \cite{DengYin2013J,HongLuo2013,ShiLing2013J,ChangTSP15_CADMM,JakoveticTAC15}, the synchronous distributed ADMM can converge linearly under similar conditions, our analysis results imply that the desirable linear convergence property of the ADMM can be preserved under the asynchronous protocol.
To the best of our knowledge, our results are novel, and are by no means extensions of
the existing analyses \cite{DengYin2013J,HongLuo2013,ShiLing2013J,ChangTSP15_CADMM,JakoveticTAC15} for synchronous ADMM. Secondly,
we present extensive numerical results to demonstrate the time efficiency of the AD-ADMM over its synchronous counterpart. In particular, we consider a large-scale LR problem and implement the AD-ADMM on a high-performance computer cluster. The presented numerical results show that the AD-ADMM significantly reduces the practical running time of distributed optimization.

{\bf Synopsis:} Section \ref{sec: async admm} reviews the AD-ADMM in \cite{ChangAsyncadmm15_p1}.
The linear convergence analysis is presented in Section \ref{sec: linear conv rate} and the proofs are presented in Section \ref{sec: proof of thm1}. Numerical results are given in Section \ref{sec: simulations} and conclusions are drawn in Section \ref{sec: conclusions}.

%
%{\textit{\bf Notations:} %Column vectors and matrices are written in boldfaced lowercase and uppercase letters, e.g., $\ab$ and $\Ab$. The superscripts $(\cdot)^T$ represents the transpose.
%$\Ab\succeq \zerob$ ($\succ \zerob$) means that matrix $\Ab$ is positive semidefinite (positive definite). $\Ib_K$ is the $K \times K$ identity matrix; $\oneb_K$ is the $K$-dimensional all-one vector.
%$\|\ab\|_2$ denotes the Euclidean norm of vector $\ab$, and $\|\zb\|^2_\Ab\triangleq \zb^T\Ab\zb$ for some $\Ab\succeq \zerob$.
%Notation $\otimes$ denotes the Kronecker product. $\diag\{a_1,\ldots,a_N\}$ is a diagonal matrix with the $i$th diagonal element being $a_i$; while $\blkdiag\{\Ab_1,\ldots,\Ab_N\}$ is a block diagonal matrix with the $i$th diagonal block matrix being $\Ab_i$. $\lambda_{\max}(\Ab)$ and $\lambda_{\min}(\Ab)$ denote the maximum and minimum eigenvalues of matrix $\Ab$, respectively.}

\vspace{-0.4cm}
\section{Asynchronous Distributed ADMM}\label{sec: async admm}
\vspace{-0.0cm}

In this section, we review the AD-ADMM proposed in \cite{ChangAsyncadmm15_p1}. The distributed ADMM \cite[Section 7.1.1]{BoydADMM11} is derived based on the following consensus formulation of \eqref{eqn: original problem}:
\begin{subequations}\label{eqn: consensus problem}
\begin{align}
  \min_{\substack{\xb_0,\xb_i\in \mathbb{R}^n, \\ i=1,\ldots,N}}~ &\sum_{i=1}^N f_i(\xb_i) + h(\xb_0)
  \\
  \text{s.t.}~ & \xb_i =\xb_0~\forall i\in \Vc\triangleq \{1,\ldots,N\}.\label{eqn: consensus problem C1}
\end{align}
\end{subequations}
%where $\Vc\triangleq \{1,\ldots,N\}$.
By applying the standard ADMM \cite{BertsekasADMM} to problem \eqref{eqn: consensus problem}, one obtains
the following three simple steps: for iteration $k=0,1,\ldots,$ update
\begin{align}
\!\!\!\!\xb^{k+1}_0\! &\!=\! \arg\min_{\xb_0 \in \mathbb{R}^n}\!  \textstyle \bigg\{\!h(\xb_0)-  \xb_0^T\sum_{i=1}^N \lambdab_i^{k} %\notag \\
%&\textstyle~~~~~~~~~~~~~~~~~~~~~~~
\!+\!\frac{\rho}{2}\sum_{i=1}^N\|\xb_i^{k}-\xb_0\|^2\bigg\},\label{eqn: sync cadmm s2}
%\\
\end{align}
\begin{align}
\xb^{k+1}_i & \!=\!\arg\min_{\xb_i \in \mathbb{R}^n}~ \textstyle f_i(\xb_i)+\xb_i^T\lambdab_i^k +\frac{\rho}{2}\|\xb_i -\xb_0^{k+1}\|^2 ~~\forall i\in \Vc,
\label{eqn: sync cadmm s1}
\\
%\end{align}
%\begin{align}
\lambdab^{k+1}_i&= \lambdab^{k}_i + \rho (\xb_i^{k+1}-\xb_0^{k+1})~~\forall i\in \Vc.\label{eqn: sync cadmm s3}
\end{align}

As seen, the distributed ADMM is designed for a computing network with a star topology that consists of one master node and a set of $N$ workers (see Fig. 1 in \cite{ChangAsyncadmm15_p1}). In particular, the master is responsible for optimizing the variable $\xb_0$ by \eqref{eqn: sync cadmm s2}, while each worker $i$, $i\in \Vc$, takes charge of optimizing variables $\xb_i$ and $\lambdab_i$ by \eqref{eqn: sync cadmm s1} and \eqref{eqn: sync cadmm s3}, respectively.
Once the master updates $\xb_0$, it broadcasts $\xb_0$ to the workers; each worker $i$ then updates $(\xb_i,\lambdab_i)$ based on the received $\xb_0$, and sends the new $(\xb_i,\lambdab_i)$ to the master.
Through such iterative variable update and message exchange, problem \eqref{eqn: consensus problem} is solved in a fully parallel and distributed fashion.

However, to implement \eqref{eqn: sync cadmm s2}-\eqref{eqn: sync cadmm s3}, the master and the workers have to be synchronized with each other.
Specifically, according to \eqref{eqn: sync cadmm s2}, the master proceeds to update $\xb_0$ only if it has received update-to-date $(\xb_i,\lambdab_i)$ from all the workers.
This implies that the optimization speed would be determined by the slowest worker in the network.
This is in particular the case in a heterogeneous network where the workers
experience different computation and communication delays, in which case the master and speedy workers would idle most of the time.

The distributed ADMM has been extended to an asynchronous network in \cite{ChangAsyncadmm15_p1,Zhang14ACADMM}. In the AD-ADMM, the master does not wait for all the workers, but updates the variable $\xb_0$ as long as it receives variable information from a partial set of workers instead. This would greatly reduce the waiting time of the master, and improve the overall time efficiency of distributed optimization. %However, the master inevitably uses delayed information for updating $\xb_0$.
The AD-ADMM is presented in Algorithm \ref{table: async cadmm s1 master}, which includes the algorithmic steps of the master and those of the workers.
Here, we denote $k$ as the iteration number of the master (i.e., the number of times for which the master updates $\xb_0$), and assume that, at each iteration $k$, the master receives variable information from workers in the set $\Ac_k \subseteq \Vc\triangleq \{1,\ldots,N\}$. Worker $i$ is said to be ``arrived" at iteration $k$ if $i\in \Ac_k$ and unarrived otherwise. Notation $\Ac_k^c$ denotes the complementary set of $\Ac_k$, i.e., $\Ac_k\cap \Ac_k^c =\emptyset$ and $\Ac_k\cup \Ac_k^c=\Vc$. Moreover, variables $d_i$'s are used to count the numbers of delayed iterations of the workers.
The variables $\rho$ and $\gamma$ are two penalty parameters.

In the AD-ADMM, the master inevitably uses delayed and old variable information for updating $\xb_0$.
As shown in step 4 of Algorithm of the Master, to ensure the used variable information not too stale, the master
would wait until it receives the update-to-date $(\xb_i,\lambdab_i)$ from all the workers that have $d_i\geq \tau-1$, if any (so all the workers $i\in \Ac_k^c$ must have $d_i<\tau-1$). This condition guarantees that the variable information is at most $\tau$ iterations old, and is known as the partially asynchronous model \cite{BertsekasADMM}:

\begin{Assumption}\label{assumption bounded delay} {\rm (Bounded delay)}
Let $\tau\geq 1$ be a maximum tolerable delay. For all $i\in \Vc$ and iteration $k$, it must be that $i\in \Ac_k \cup \Ac_{k-1}\cdots \cup \Ac_{k-\tau+1}$.
\end{Assumption}

In \cite[Theorem 1]{ChangAsyncadmm15_p1}, we have shown that under Assumption \ref{assumption bounded delay}, some smoothness conditions on the cost functions $f_i$'s (see \cite[Assumption 2]{ChangAsyncadmm15_p1}) and for sufficiently large $\rho$ and $\gamma$, the AD-ADMM in Algorithm \ref{table: async cadmm s1 master} is provably convergent to the set of KKT points of problem \eqref{eqn: consensus problem}. Notably, this convergence property holds even for non-convex $f_i$'s.
In the next section, we focus on convex $f_i$'s, and further characterize the linear convergence conditions of the AD-ADMM.

\begin{algorithm}[h!] \label{table: async cadmm s1}
\caption{Asynchronous Distributed ADMM for \eqref{eqn: consensus problem}.}
\begin{algorithmic}[1]\label{table: async cadmm s1 master}
\STATE {\bf \underline{Algorithm of the Master:}}
\STATE {\bf Given} initial variable
$\xb^{0}$ and broadcast it to the workers. Set $k=0$ and $d_1=\cdots=d_N=0;$
\REPEAT
\STATE  {\bf wait} until receiving $\{\hat\xb_i,\hat\lambdab_i\}_{i\in \Ac_k}$ from workers $i\in \Ac_k$ and
that $d_i <\tau-1$ $\forall i \in \Ac_k^c$.
\STATE {\bf update}
\begin{align}&\xb^{k+1}_i =\bigg\{\begin{array}{ll}
\hat\xb_i  & \forall i\in \Ac_k \\
\xb^{k}_i & \forall i\in \Ac_k^c
\end{array},  \label{eqn: async cadmm s1 xii} \\
&\lambdab^{k+1}_i =\bigg\{\begin{array}{ll}
\hat\lambdab_i  & \forall i\in \Ac_k \\
\lambdab^{k}_i & \forall i\in \Ac_k^c
\end{array}, \label{eqn: async cadmm s1 lambdaii} \\
 &d_i =\bigg\{\begin{array}{ll}
0  & \forall i\in \Ac_k \\
d_i+1 & \forall i\in \Ac_k^c
\end{array}, \\
&\xb^{k+1}_0 \! =\!\arg\min_{\xb_0 \in \mathbb{R}^n} \textstyle  \bigg\{h(\xb_0)-  \xb_0^T\sum_{i=1}^N \lambdab_i^{k+1} \notag \\
&~\textstyle+\frac{\rho}{2}\sum_{i=1}^N\|\xb_i^{k+1}-\xb_0\|^2+\frac{\gamma}{2}\|\xb_0-\xb_0^k\|^2\bigg\},\label{eqn: async cadmm s1 x0}
\end{align}
\STATE {\bf broadcast} $\xb^{k+1}_0$ to the workers in $\Ac_k$.
\STATE {\bf set} $k\leftarrow k+1.$
\UNTIL {a predefined stopping criterion is satisfied.}%
\end{algorithmic}
%\end{algorithm

%\begin{algorithm}[h]
%\caption{Asynchronous Consensus ADMM (Scheme 1): The $i$th Worker }
\begin{algorithmic}[1]\label{table: async cadmm s1 worker}

\STATE {\bf \underline{Algorithm of the $i$th Worker:}}
\STATE {\bf Given} initial $\lambdab^{0}$ and set ${k_i}=0.$
%\STATE {\bf Set} $k=1.$
\REPEAT
\STATE  {\bf wait} until receiving $\hat\xb_0$ from the master node.
\STATE {\bf update}
\begin{align}
\xb_i^{{k_i}+1} &=\arg\min_{\xb_i\in \mathbb{R}^n} \textstyle f_i(\xb_i)+\xb_i^T\lambdab_i^{k_i}+\frac{\rho}{2}\|\xb_i -\hat\xb_0\|^2,
\label{eqn: async cadmm s1 xi}\\
 \lambdab_i^{{k_i}+1}&= \lambdab^{{k_i}}_i + \rho (\xb_i^{{k_i}+1}-\hat \xb_0).\label{eqn: async cadmm s1 lambda}
\end{align}
\STATE {\bf send} $(\xb_i^{{k_i}+1},\lambdab_i^{{k_i}+1})$ to the master node.
\STATE {\bf set} ${k_i}\leftarrow {k_i}+1.$
\UNTIL {a predefined stopping criterion is satisfied.}%
\end{algorithmic}
\end{algorithm}

\section{Linear Convergence Rate Analysis}\label{sec: linear conv rate}

%\subsection{Assumption and Main Result}
In this section, we show that the AD-ADMM can achieve linear convergence for some structured convex functions.
%\subsection{Linear Convergence Conditions}
We first make the following convex assumption on problem \eqref{eqn: original problem} (or equivalently, problem \eqref{eqn: consensus problem}).
\begin{Assumption} \label{assumption obj s1}
Each function $f_i$ is a proper closed convex function and is continuously differentiable; each gradient $\nabla f_i$ is Lipschitz continuous with a Lipschitz constant $L>0$; the function $h$ is convex (not necessarily smooth).
Moreover, problem \eqref{eqn: original problem} is bounded below, i.e., $F^\star >-\infty$ where $F^\star$ denotes the optimal objective value of problem \eqref{eqn: original problem}.
\end{Assumption}

Assumption \ref{assumption obj s1} is the same as \cite[Assumption 2]{ChangAsyncadmm15_p1}, except that $f_i$'s are assumed convex here. Given this convex property, it is well known that the augmented Lagrangian function, i.e.,
\begin{align}\label{eqn: Lc}
  \Lc_{\rho}(\xb^k,\xb_0^k, \lambdab^k)=
  \sum_{i=1}^N f_i(\xb_i^k) &+ h(\xb_0^k)+\sum_{i=1}^N (\lambdab_i^k)^T(\xb_i^k -\xb_0^k) \notag \\
  &+\frac{\rho}{2}\sum_{i=1}^N\|\xb_i^k -\xb_0^k\|^2,
\end{align}
would converge to $F^\star$ whenever the iterates $(\{\xb_i^k\}_{i=1}^N,\xb_0^k,$ $ \{\lambdab_i^k\}_{i=1}^N)$ approaches the optimal solution of problem \eqref{eqn: consensus problem}. Therefore, our analysis is based on characterizing how
$\Lc_{\rho}(\xb^k,\xb_0^k, \lambdab^k)$ can converge to $F^\star$ linearly. Let us define
\begin{align}
  \triangle_{k}\triangleq \Lc_{\rho}(\xb^k,\xb_0^k, \lambdab^k)-F^\star.
\end{align}
It has been shown in \cite[Lemma 3]{ChangAsyncadmm15_p1} that $\triangle_{k}\geq 0$ for all $k$ as long as $\rho\geq L$.
%{\red[[may need to briefly discuss the fact that the gap is always positive during the iteration. Otherwise people may question the validity of using this as a measure. ]]}

In the ensuing analysis, we consider two types of structured convex cost functions, respectively described in the following two assumptions.

\begin{Assumption}\label{assumption scf & osc}
  For all $i\in \Vc,$ each function $f_i$ is strongly convex with modulus $\sigma^2>0$.
\end{Assumption}

\begin{Assumption}\label{assumption scf & osc2}
  Each function $f_i(\xb)=g_i(\Ab_i\xb)$, $\forall i\in \Vc,$ where $g_i:\mathbb{R}^m\to\mathbb{R}$ is a strongly convex function with modulus $\sigma^2>0$ and $\Ab_i\in \mathbb{R}^{m\times n}$ is a nonzero matrix with arbitrary rank. Moreover, $h(\xb)=0$.
\end{Assumption}

Note that in Assumption \ref{assumption scf & osc2} matrix $\Ab_i$ can have an arbitrary rank, so $f_i(\xb)$ is not necessarily strongly convex with respect to $\xb$. %Therefore, %Assumption \ref{assumption scf & osc2} is weaker than Assumption \ref{assumption scf & osc}, although the non-smooth function $h$ has to be absent.
Interestingly, such structured cost function appears in many machine learning problems, for example, the least squared problem and the logistic regression problem \cite{Hastie2001Book}.

Let us first consider the strongly convex case. Under Assumption \ref{assumption scf & osc}, the linear convergence conditions of the AD-ADMM are given by the following theorem.

\begin{Theorem} \label{thm: linear conv of scheme 1}Suppose that Assumptions \ref{assumption bounded delay}, \ref{assumption obj s1} and \ref{assumption scf & osc} hold true. Moreover, assume that there exists a constant $S\in [1, N]$ such that $|\Ac_k|<S$ for all $k$ and that
\begin{align}\label{eqn: linear cond of conv scheme1}
    &\rho\geq \max\bigg\{\frac{(1+L^2)+\sqrt{(1+L^2)^2+8L^2\alpha(\tau)}}{2},\sigma^2+ \frac{1}{8N}\bigg\},
    \\
    &\gamma \geq \max\bigg\{\beta(\rho,\tau)-\frac{N\rho}{2}+1, 8N(\rho-\sigma^2)\bigg\}, \label{eqn: linear cond of conv scheme1 2}
\end{align}
where $\alpha(\tau)\triangleq 1+\frac{2+2^{\tau}(\tau-1)}{1+8N\sigma^2}$ and
$\beta(\rho,\tau)\triangleq   2(\tau-1)[(\frac{(1+\rho^2)S+S/N}{2})({2^{\tau-1}-1})+{({4^{\tau-1}-1})}]$. Then, the iterates generated by \eqref{eqn: async cadmm s1 xii}, \eqref{eqn: async cadmm s1 lambdaii} and \eqref{eqn: async cadmm s1 x0} satisfy
\begin{align}\label{eqn: linear conv rate}
  0\leq
  \triangle_{k+1}\leq
  \bigg(\frac{1}{1+\frac{1}{\delta\gamma }}\bigg)^{k+1}\triangle_0,
\end{align} where $\delta$ is a constant satisfying
\begin{align}\label{eqn: linear cond of conv scheme1 3}
  %&\frac{ {\rho N}\bigg[(1+\frac{1}{\delta})^2-1\bigg] +\frac{\gamma}{\delta}}{N(1+\frac{1}{\delta})^2} \leq \mu,\\
  &\delta \geq \max\bigg\{1,\frac{\rho N+\gamma}{\sigma^2 N}-1\bigg\}.
\end{align}
%Here, $\mu=\sigma^2$ for the case of Assumption \ref{assumption scf & osc}(a) and $\mu=\sigma^2/c$, where $c>0$ is a constant, for the case of Assumption \ref{assumption scf & osc}(b).
\end{Theorem}

Theorem \ref{thm: linear conv of scheme 1} asserts that, for problem \eqref{eqn: consensus problem} with strongly convex $f_i$'s, the augmented Lagrange function can decrease linearly to zero, as long as $\rho$ and $\gamma$ are large enough (exponentially increasing with $\tau$).
Equation \eqref{eqn: linear conv rate} also implies that the linear rate would decrease with the delay $\tau$ and the number of workers in the worst case. %Interestingly, when either $\tau$ is small or $N/S$ is large, one can further have the following result:

%\begin{Corollary} \label{coro: linear conv of scheme 1}Suppose that Assumptions \ref{assumption bounded delay}, \ref{assumption obj s1} and \ref{assumption scf & osc} hold true and that $|\Ac_k|<S$ for all $k$ for some constant $S\in [1, N]$. Moreover, assume that
%$\gamma=0$, $(\tau-1)(2^{\tau-1}-1)< \frac{N}{2S}\epsilon$ for some $\epsilon \in (0,1)$ and %$N>4S^2$,
%\begin{align}
%    \label{eqn: linear cond of conv scheme1 gamma0}
%    &\rho\geq \max\bigg\{\frac{L^2+\sqrt{L^4+8L^2(1-\epsilon)\alpha(N)}}{2},\sigma^2+ \frac{1}{4N},
%    \notag \\
%    & \frac{(1+\frac{2}{N})S(\tau-1)(2^{\tau-1}-1) + 4(\tau-1)(4^{\tau-1}-1)+1}{N/2 - S(\tau-1)(2^{\tau-1}-1)/\epsilon}
%    \bigg\},
%\end{align}
%where $\alpha(N)\triangleq 1+\frac{2+2^\tau(\tau-1)}{1+4N\sigma^2}$. Then, the iterates generated by \eqref{eqn: async cadmm s1 xii}, \eqref{eqn: async cadmm s1 lambdaii} and \eqref{eqn: async cadmm s1 x0} satisfy
%\begin{align}\label{eqn: linear conv rate gamma0}
%  0\leq
%  \triangle_{k+1}\leq
%  \bigg(\frac{1}{1+\frac{1}{4(\rho-\sigma^2)N\delta }}\bigg)^{k+1}\triangle_0,
%\end{align} where $\delta$ is a constant satisfying $\delta\geq \max\{\rho/\sigma^2-1,1\}$.
%\end{Corollary}
%
%Corollary \ref{coro: linear conv of scheme 1} indicates that when $\tau = \mathcal{O}(\log_2 \frac{N}{S})$, the AD-ADMM can converge linearly even with $\gamma=0$. Moreover, when $N/S \to \infty$, the right hand side of \eqref{eqn: linear cond of conv scheme1 gamma0} tends to a constant, which implies that a delay-independent $\rho$ may be sufficient for achieving linear convergence.

Analogous to Theorem \ref{thm: linear conv of scheme 1}, the following theorem shows that the AD-ADMM can achieve linear convergence under Assumption \ref{assumption scf & osc2}. %{\color{red}[[why this theorem is stated so vaguely?]]}
\begin{Theorem} \label{thm: linear conv of scheme 1 2}
Suppose that Assumptions \ref{assumption bounded delay}, \ref{assumption obj s1} and \ref{assumption scf & osc2} hold true.
%, and assume that $\rho$ and $\gamma$ are sufficiently large. Then
%$\triangle_{k+1}$ converges to zero in a linear rate. Moreover, if $\tau = \mathcal{O}(\log_2 N)$ and $\rho$ is sufficiently %large, then
%$\triangle_{k+1}$ converges to zero linearly.
%the iterates generated by \eqref{eqn: async cadmm s1 xi}, \eqref{eqn: async cadmm s1 lambdai} and \eqref{eqn: async cadmm s1 x0} satisfy
%\begin{align}\label{eqn: linear conv rate 2}
%  0\leq
%  \triangle_{k+1}\leq
%  \bigg(\frac{1}{1+\frac{1}{\delta\gamma }}\bigg)^{k+1}\triangle_0,
%\end{align} where $\delta$ is a constant satisfying
%\begin{align}\label{eqn: linear cond of conv scheme1 3}
  %&\frac{ {\rho N}\bigg[(1+\frac{1}{\delta})^2-1\bigg] +\frac{\gamma}{\delta}}{N(1+\frac{1}{\delta})^2} \leq \mu,\\
%  &\delta \geq \max\bigg\{1,\frac{\rho N+\gamma}{\sigma^2 N/c}-1\bigg\}.
%\end{align}
%for some constant $c>0$.
Moreover, assume that there exists a constant $S\in [1, N]$ such that $|\Ac_k|<S$ for all $k$ and that
\begin{align*}%\label{eqn: linear cond of conv scheme1}
    &\rho\geq \max\bigg\{\frac{(1+L^2)+\sqrt{(1+L^2)^2+8L^2\alpha(\tau)}}{2},\sigma^2+ \frac{1}{8N}\bigg\},
    \\
    &\gamma \geq \max\bigg\{\beta(\rho,\tau)-\frac{N\rho}{2}+1, 8N(\rho-\sigma^2/c)+4N\sigma^2\bigg\}, %\label{eqn: linear cond of conv scheme1 2}
\end{align*}
%where $\alpha(\tau)\triangleq 1+\frac{2+2^{\tau}(\tau-1)}{1+8N\sigma^2}$ and
%$\beta(\rho,\tau)\triangleq   2(\tau-1)[(\frac{(1+\rho^2)S+S/N}{2})({2^{\tau-1}-1})+{({4^{\tau-1}-1})}]$.
for some constant $c>0$. Then, the iterates generated by \eqref{eqn: async cadmm s1 xii}, \eqref{eqn: async cadmm s1 lambdaii} and \eqref{eqn: async cadmm s1 x0} satisfy \eqref{eqn: linear conv rate}
%\begin{align}\label{eqn: linear conv rate}
%  0\leq
%  \triangle_{k+1}\leq
%  \bigg(\frac{1}{1+\frac{1}{\delta\gamma }}\bigg)^{k+1}\triangle_0,
%\end{align}
with $\delta$ satisfying
\begin{align}%\label{eqn: linear cond of conv scheme1 32}
  %&\frac{ {\rho N}\bigg[(1+\frac{1}{\delta})^2-1\bigg] +\frac{\gamma}{\delta}}{N(1+\frac{1}{\delta})^2} \leq \mu,\\
  &\delta \geq \max\bigg\{1,\frac{\rho N+\gamma}{N\sigma^2/c}-1\bigg\}. \notag
\end{align}
\end{Theorem}

%One can also show a similar result as Corollary \ref{coro: linear conv of scheme 1} that the AD-ADMM converges linearly with $\gamma=0$ if $\tau = \mathcal{O}(\log_2 \frac{N}{S})$ and $\rho$ is sufficiently large.

Since it has been known that the (synchronous) distributed ADMM \cite{DengYin2013J,HongLuo2013,ShiLing2013J,ChangTSP15_CADMM,JakoveticTAC15} can converge linearly given the same structured cost functions in Assumption \ref{assumption scf & osc} and Assumption \ref{assumption scf & osc2}, the convergence results presented above demonstrate that the linear convergence property can be preserved in the asynchronous network.
%Equation \eqref{eqn: linear conv rate} further suggests that the linear rate can decrease with the maximal tolerable delay $\tau$ and the number of workers. These aspects will be further examined via simulations in Section \ref{sec: simulations}.
%Note that \eqref{eqn: linear cond of conv scheme1} and \eqref{eqn: linear cond of conv scheme1 2} are sufficient conditions only. In practice,
%the AD-ADMM could still exhibit a linear convergence rate without exactly satisfying these conditions.
%To the best of our knowledge, Theorem \ref{thm: linear conv of scheme 1} and Theorem \ref{thm: linear conv of scheme 1 2} are the first set of theoretical results on the linear convergence of the AD-ADMM, and have never been discussed in the literature.
We remark that \eqref{eqn: linear cond of conv scheme1} and \eqref{eqn: linear cond of conv scheme1 2} %and \eqref{eqn: linear cond of conv scheme1 gamma0} 
are sufficient conditions only. In practice,
the AD-ADMM could still exhibit a linear convergence rate without exactly satisfying these conditions.

The proofs of  Theorem \ref{thm: linear conv of scheme 1} %, Corollary \ref{coro: linear conv of scheme 1} 
and Theorem \ref{thm: linear conv of scheme 1 2} are presented in the next section. The readers who are more interested in the numerical performance of the AD-ADMM may jump to Section \ref{sec: simulations}.

%\begin{Remark}\rm
%Note that \eqref{eqn: linear cond of conv scheme1} and \eqref{eqn: linear cond of conv scheme1 2} are sufficient conditions only. In practice,
%the AD-ADMM could still exhibit a linear convergence rate without exactly satisfying these conditions.
%Indeed, as we have shown in Figures 3(a) and 3(c) in the companion paper \cite{ChangAsyncadmm15_p1} and in the next section, the AD-ADMM may converge linearly for some problems even with $\gamma=0$%\footnote{In practice, it is fine to set $\gamma>0$. However, the performance may be similar to that with $\gamma=0$. This is because, as one can see from \eqref{eqn: async cadmm s1 x0}, both parameter $\rho$ and $\gamma$ can control the step size of $\xb_0$. Therefore, in many cases, it is sufficient to leave $\gamma$ equal to zero. }
%. %The numerical results presented in \cite{Zhang14ACADMM} also show that the AD-ADMM
%While this shows the gap between theoretical analysis and practice, we should emphasize that, like \cite{DengYin2013J,HongLuo2013,ShiLing2013J,ChangTSP15_CADMM,JakoveticTAC15}, our analysis results provide not only theoretical performance guarantees but also useful insights on how the network parameters (e.g., $\tau$, $N$ and penalty parameters $\rho$) affect the convergence behavior of the AD-ADMM.
%\end{Remark}

\section{Proofs of Theorems}\label{sec: proof of thm1}

%Let us present the proof of Theorem \ref{thm: sublinear conv rate} in this section.

\subsection{Preliminaries and Key Lemmas}

Let us present some basic inequalities that will be used frequently in the ensuing analysis and key lemmas for proving
Theorem \ref{thm: linear conv of scheme 1} and Theorem \ref{thm: linear conv of scheme 1 2}.

We will frequently use the following inequality due to Jensen's inequality: for any $\ab_i$, $i=1,\ldots,M$,
\begin{align}\label{eqn: 12norm}
  \textstyle \|\sum_{i=1}^M \ab_i\|^2\leq M \sum_{i=1}^M\|\ab_i\|^2.
\end{align}
Moreover, for any $\ab$, $\bb$ and $\delta>0$,
\begin{align}
  %\ab^T\bb &\leq \frac{\delta}{2}\|\ab\|^2 + \frac{1}{2\delta}\|\bb\|^2, \label{eqn: young ineq}\\
  \|\ab+\bb\|^2 & \leq (1+\delta) \|\ab\|^2 +(1+\frac{1}{\delta}) \|\bb\|^2. \label{eqn: young ineq2}
\end{align}
The equality is also known to be true: for any vectors $\ab$, $\bb$, $\cb$ and $\db$,
\begin{align}\label{eqn: identity}
 &(\ab-\bb)^T(\cb-\db) = \frac{1}{2}\|\ab-\db\|^2-\frac{1}{2}\|\ab-\cb\|^2
 \notag \\
 &~~~~~~~+ \frac{1}{2}\|\bb-\cb\|^2
 %\notag \\
 -\frac{1}{2}\|\bb-\db\|^2.
\end{align}
%The following inequality due to Jensen's inequality is also used: for any $\ab_i$, $i=1,\ldots,M$,
%\begin{align}\label{eqn: 12norm}
%  \textstyle \|\sum_{i=1}^M \ab_i\|^2\leq M \sum_{i=1}^M\|\ab_i\|^2.
%\end{align}

We follow \cite[Algorithm 3]{ChangAsyncadmm15_p1} to write Algorithm \ref{table: async cadmm s1 master} from the master's point of view as follows:  %Specifically, let $k$ denotes the iteration number of the master (i.e., the number of times for which the master updates $\xb_0$).
%Specifically, one can write Algorithm \ref{table: async cadmm s1 master} equivalently as follows
\begin{align}
\xb^{k+1}_i &\!=\!\left\{\!\!\begin{array}{ll}
\arg{\displaystyle \min_{\xb_i\in \mathbb{R}^n}} \textstyle \bigg\{f_i(\xb_i)+\xb_i^T\lambdab^{k}_i +\frac{\rho}{2}\|\xb_i -\xb_0^{\bar k_i +1}\|^2\bigg\}, & \forall i\in \Ac_k
\\
\xb^{k}_i & \forall i\in \Ac_k^c
\end{array}\right.,\label{eqn: async cadmm s1 xi equi2}
\\
%\end{align}
%\begin{align}
\lambdab^{k+1}_i &=\bigg\{\begin{array}{ll}
\lambdab^{k}_i + \rho (\xb_i^{k+1}-\xb_0^{\bar k_i +1})  & \forall i\in \Ac_k \\
\lambdab^{k}_i & \forall i\in \Ac_k^c
\end{array}, \label{eqn: async cadmm s1 lambda equi2} \\
\xb^{k+1}_0 \! &=\!\arg\min_{\xb_0 \in \mathbb{R}^n} \textstyle  \bigg\{h(\xb_0)-  \xb_0^T\sum_{i=1}^N \lambdab_i^{k+1} \notag \\
&~\textstyle+\frac{\rho}{2}\sum_{i=1}^N\|\xb_i^{k+1}-\xb_0\|^2+\frac{\gamma}{2}\|\xb_0-\xb_0^k\|^2\bigg\}.\label{eqn: async cadmm s1 x0 equi2}
\end{align}
Here, index $\bar k_i$ in \eqref{eqn: async cadmm s1 xi equi2} and \eqref{eqn: async cadmm s1 lambda equi2} represents the last iteration number before iteration $k$ for which worker $i\in \Ac_k$ is arrived, i.e., $i\in \Ac_{\bar k_i}$.
Under Assumption 1, it must hold
\begin{align}\label{eqn: cond on bar ki}
    k-\tau \leq  \bar k_i <k~~\forall k.
\end{align}
Furthermore, for workers $i\in \Ac_k^c$, let us denote $\widetilde k_i$
as the last iteration number before iteration $k$ for which worker $i$ is arrived, i.e., $i\in \Ac_{\widetilde k_i}$. Then, under Assumption 1, it must hold
\begin{align}\label{eqn: cond on bar ki2}
    k-\tau<\widetilde k_i<k~~\forall k.
\end{align}
In addition, denote $\widehat k_i$ ($\widetilde k_i-\tau \leq \widehat k_i <\widetilde k_i$) as the last iteration number before iteration $\widetilde k_i$ for which worker $i\in \Ac_{\widetilde k_i}$ is arrived, i.e.,
$i\in \Ac_{\widehat k_i}$. Then by \eqref{eqn: async cadmm s1 xi equi2} and \eqref{eqn: async cadmm s1 lambda equi2}, for all workers $i\in \Ac_k^c$, we must have \begin{align}\label{eqn: unchaged x}
&\xb^{\widetilde k_i+1}_i=\xb^{\widetilde k_i+2}_i=\cdots=\xb^{k}_i=\xb^{k+1}_i,\\
&\lambdab^{\widetilde k_i+1}_i=\lambdab^{\widetilde k_i+2}_i=\cdots=\lambdab^{k}_i=\lambdab^{k+1}_i, \label{eqn: unchaged lambda1}
%\\
%& \xb^{\widehat k_i+1}_i=\xb^{\widehat k_i+2}_i=\cdots=\xb^{\widetilde k_i}_i, \label{eqn: unchaged x2} \\
%& \lambdab^{\widehat k_i+1}_i=\lambdab^{\widehat k_i+2}_i=\cdots=\lambdab^{\widetilde k_i}_i.\label{eqn: unchaged lambda2}
\end{align}
Since $i\in \Ac_{\widetilde k_i}$ for all $i\in \Ac_k^c$ and by \eqref{eqn: unchaged x}-\eqref{eqn: unchaged lambda1}, we can equivalently write \eqref{eqn: async cadmm s1 xi equi2} and \eqref{eqn: async cadmm s1 lambda equi2} for all $i\in \Ac_k^c$ as
\begin{align}
\xb^{k+1}_i&=\xb^{\widetilde k_i+1}_i \notag\\
%&=\arg~{\displaystyle \min_{\xb_i}}~ \textstyle f_i(\xb_i)+\xb_i^T\lambdab^{\widehat k_i+1}_i +\frac{\rho}{2}\|\xb_i -\xb_0^{\widehat k_i+1}\|^2 \notag \\
&=\arg~{\displaystyle \min_{\xb_i}}~ \textstyle f_i(\xb_i)+\xb_i^T\lambdab^{\widetilde k_i }_i +\frac{\rho}{2}\|\xb_i -\xb_0^{\widehat k_i+1}\|^2, \label{eqn: async cadmm s1 xi skc equi} \\
\lambdab^{k+1}_i&
=\lambdab^{\widetilde k_i+1}_i  = \lambdab^{\widetilde k_i}_i + \rho (\xb_i^{\widetilde k_i+1}-\xb_0^{\widehat k_i +1}) \notag \\
&~~~~~~~~~~~= \lambdab^{\widetilde k_i}_i + \rho (\xb_i^{k+1}-\xb_0^{\widehat k_i +1}). \label{eqn: async cadmm s1 lambda skc equi}
\end{align}

Based on these notations, we have shown in \cite[Eqn. (33)]{ChangAsyncadmm15_p1} that the following lemma is true.

\begin{Lemma} \label{lemma: Lc progress} Suppose that Assumption \ref{assumption obj s1} holds and $\rho\geq L$. Then, for all $k=0,1,\ldots,$
\begin{align}
0\leq &\triangle_{k+1} \leq
\triangle_{k} + \bigg(\frac{1+\rho/\epsilon}{2}\bigg)\!\!\sum_{i\in \Ac_k} \|\xb_0^{k}-\xb_0^{\bar k_i+1}\|^2 \notag \\
&
-\!
\bigg(\frac{2\gamma+N\rho}{2}\bigg)\|\xb_0^{k+1}-\xb_0^{k}\|^2
+ \bigg(\frac{L^2+(\epsilon-1)\rho}{2}+\frac{L^2}{\rho}\bigg)\sum_{i\in \Ac_k} \|\xb_i^{k+1}-\xb_i^{k}\|^2, \label{lemma: Lc progress eqn0}
\end{align}
where $\epsilon\in (0,1)$ is a constant.
\end{Lemma}

In particular, \eqref{lemma: Lc progress eqn0} is the same as \cite[Eqn. (33)]{ChangAsyncadmm15_p1} except that here we have assumed convex $f_i$'s. Lemma \ref{lemma: Lc progress} shows how the gap between the augmented Lagrangian
function $\Lc_{\rho}(\xb^{k+1},\xb_0^{k+1}, \lambdab^{k+1})$ and the optimal objective value $F^\star$ evolves with the iteration number $k$.
Notice that it follows from \cite[Lemma 3]{ChangAsyncadmm15_p1} that $\triangle_{k+1} \geq 0$ for all $k$ if $\rho\geq L$.
As will be seen shortly, Lemma \ref{lemma: Lc progress} is crucial in the linear convergence analysis.

%Since problem \eqref{eqn: consensus problem} is a consensus optimization problem, the next lemma bounds the consensus error $\sum_{i=1}^N\|\xb_i^{k+1}\!-\!\xb_0^{k+1}\|^2$ by successive differences of variables $\xb_i^k$, $i\in \Vc$, and $\xb_0^k$.
%
%\begin{Lemma} \label{lemma: consensus error} Under Assumption \ref{assumption obj s1}, it holds that
%\begin{align}\label{eqn: consensus error bound}
%\!\!\!\!\!  &\sum_{i=1}^N\|\xb_i^{k+1}\!-\!\xb_0^{k+1}\|^2\leq \frac{2L^2}{\rho^2}\sum_{i\in \Ac_k}\|\xb_i^{k+1}\!-\xb_i^{k}\|^2 \notag \\
%\!\!\!\!\!  &
%   ~~~+\frac{2L^2}{\rho^2}\sum_{i\in \Ac_k^c}\|\xb_i^{\widetilde k_i+1}\!-\xb_i^{\widetilde k_i}\|^2
%   +4\sum_{i\in \Ac_k}\|\xb_0^{\bar k_i+1}-\!\xb_0^{k}\|^2\notag \\
%   &~~~+4\sum_{i\in \Ac_k^c}\|\xb_0^{\widehat k_i+1}-\!\xb_0^{k}\|^2+4N\|\xb^{k+1}_0-\xb_0^k\|^2.
%\end{align}
%\end{Lemma}
%
%The proof is presented in Section \ref{appx: proof of lemma: consensus error}.
Similar to \cite[Lemma 3]{ChangAsyncadmm15_p1}, we next need to bound the error terms, e.g., $(\frac{1+\rho^2}{2})\!\!\sum_{i\in \Ac_k} \|\xb_0^{k}-\xb_0^{\bar k_i+1}\|^2$ in \eqref{lemma: Lc progress eqn0}, which is caused by asynchrony of the network. Here, we present a more general result for the latter analysis.

\begin{Lemma}\label{lemma: asyn error bound} Let $\eta>0$ and $j- \nu \leq j_i <j$ where $\nu \in \mathbb{Z}_{++}$,
$j_i \in \mathbb{Z}_+$ and $j\in \{0,1,\ldots,k\}$. Moreover, let $\Nc_j\subset \Vc$ be any index subset satisfying
$|\Nc_j|\leq \bar N$ for some constant $\bar N\in (1,N]$. Then, the following inequality holds true
\begin{align}%\label{lemma: asyn error bound 0}
%&\sum_{j=0}^k \eta^j\sum_{i\in \Nc_{j}} \|\xb_i^{j}-\xb_i^{j_i+1}\|^2 \notag \\
%&~~\leq (\nu-1)\sum_{i=1}^N\sum_{j=0}^{k-1} \eta^{j+1}\bigg(\frac{\eta^{\nu-1}-1}{\eta-1}\bigg)\|\xb_i^{j}-\xb_i^{j+1}\|^2,\\
&\sum_{j=0}^k \eta^j\sum_{i\in \Nc_{j}} \|\xb_0^{j}-\xb_0^{j_i+1}\|^2 \leq (\nu-1){\bar N}\sum_{j=0}^{k-1} \eta^{j+1}\bigg(\frac{\eta^{\nu-1}-1}{\eta-1}\bigg)\|\xb_0^{j}-\xb_0^{j+1}\|^2.
\label{lemma: asyn error bound 1}
\end{align}
\end{Lemma}
{\bf Proof:} See Appendix \ref{appx: proof of lemma: asyn error bound}. \hfill $\blacksquare$
%The proof is presented in Section \ref{appx: proof of lemma: asyn error bound}.

Now let us consider Assumption \ref{assumption scf & osc}.
For strongly convex $f_i's$, it is known that the following first-order condition holds \cite{BK:BoydV04}: $\forall \xb,\yb,$
\begin{align}\label{eqn: SC}
  f_i(\yb) \geq   f_i(\xb)+ (\nabla f_i(\xb))^T(\yb-\xb) + \frac{\sigma^2}{2} \|\yb-\xb\|^2.
\end{align}
Based on this property, we can bound $\triangle_{k+1}$ as follows.

\begin{Lemma} \label{lemma: LC upper bound} Suppose that Assumptions \ref{assumption obj s1} and \ref{assumption scf & osc} hold and $\rho\geq \sigma^2$.
If $\gamma \geq 8N(\rho-\sigma^2)$ and $\delta$ satisfies \eqref{eqn: linear cond of conv scheme1 3}, then
%\begin{align}
%  \frac{ {\rho N}\bigg[(1+\frac{1}{\delta})^2-1\bigg] +\frac{\gamma}{\delta}}{N(1+\frac{1}{\delta})^2} \leq \mu.
%\end{align}
it holds that
 \begin{align}
 &\frac{1}{\gamma \delta}\triangle_{k+1} \leq \frac{ L^2}{4\rho^2 N}\sum_{i\in \Ac_k}\|\xb_i^{k+1}-\xb_i^{k}\|^2
  \notag
   %\\
   \end{align}
   \begin{align}\label{eqn: lemma LC upper bound 0}
  &~~~+ \frac{L^2}{4\rho^2 N}\sum_{i\in \Ac_k^c}\|\xb_i^{\widetilde k_i+1}-\xb_i^{\widetilde k_i}\|^2+\frac{1}{2N}\sum_{i\in \Ac_k}\|\xb_0^{k}-\xb_0^{\bar k_i+1}\|^2
  \notag \\
  &~~~+\frac{1}{2N}\sum_{i\in \Ac_k^c}\|\xb_0^{k}-\xb_0^{\widehat k_i+1}\|^2 +\|\xb_0^{k+1}-\xb_0^{k}\|^2.
\end{align}
Instead, if $\gamma=0$ and $\delta\geq \max\{\rho/\sigma^2-1,1\}$, then it holds
 \begin{align}\label{eqn: lemma LC upper bound 0 gamma0}
 &\bigg(\frac{1}{4(\rho-\sigma^2)N \delta}\bigg)\triangle_{k+1} \leq \frac{ L^2}{2\rho^2 N}\sum_{i\in \Ac_k}\|\xb_i^{k+1}-\xb_i^{k}\|^2
  \notag
   \\
  &~~~+ \frac{L^2}{2\rho^2 N}\sum_{i\in \Ac_k^c}\|\xb_i^{\widetilde k_i+1}-\xb_i^{\widetilde k_i}\|^2+\frac{1}{N}\sum_{i\in \Ac_k}\|\xb_0^{k}-\xb_0^{\bar k_i+1}\|^2
  \notag \\
  &~~~+\frac{1}{N}\sum_{i\in \Ac_k^c}\|\xb_0^{k}-\xb_0^{\widehat k_i+1}\|^2 +\|\xb_0^{k+1}-\xb_0^{k}\|^2.
\end{align}
\end{Lemma}
{\bf Proof:} See Appendix \ref{appx: proof of lemma: LC upper bound}. \hfill $\blacksquare$

\subsection{Proof of Theorem \ref{thm: linear conv of scheme 1}}\label{subsec: proof of thm1}

We use the lemmas above to prove Theorem \ref{thm: linear conv of scheme 1}.
Denote $\eta\triangleq 1+\frac{1}{\delta\gamma }.$ By summing \eqref{lemma: Lc progress eqn0} and \eqref{eqn: lemma LC upper bound 0}, we obtain
\begin{align}\label{eqn: proof of linear conv 1}
 &\triangle_{k+1} \leq
 \frac{1}{\eta}\triangle_{k} + \frac{1}{ \eta}\bigg[\bigg(\frac{L^2+(\epsilon-1)\rho+\frac{ L^2}{2\rho^2 N}}{2}+\frac{L^2}{\rho}\bigg)\sum_{i=1}^N \|\xb_i^{k+1}-\xb_i^{k}\|^2
\notag
\\
% \end{align}
% \begin{align}
&~~~~~~-
\bigg(\frac{2\gamma+N\rho}{2}-1\bigg)\|\xb_0^{k+1}-\xb_0^{k}\|^2 +\frac{1}{2N}\sum_{i\in \Ac_k^c}\|\xb_0^{k}-\xb_0^{\widehat k_i+1}\|^2 \notag \\
&~~~~~~+\frac{ L^2}{4\rho^2 N}\sum_{i\in \Ac_k^c}\|\xb_i^{\widetilde k_i+1}-\xb_i^{\widetilde k_i}\|_2^2 
+\bigg(\frac{1+\rho/\epsilon}{2}+\frac{1}{2N}\bigg)\sum_{i\in \Ac_k} \|\xb_0^{k}-\xb_0^{\bar k_i+1}\|^2\bigg].
\end{align}
Here, we have used the fact of $\sum_{i\in \Ac_k} \|\xb_i^{k+1}-\xb_i^{k}\|^2=\sum_{i=1}^N \|\xb_i^{k+1}-\xb_i^{k}\|^2$
as $\xb_i^{k+1}=\xb_i^k~\forall i\in \Ac_k^c$.
By taking the telescoping sum of \eqref{eqn: proof of linear conv 1}, we further obtain
%\eqref{eqn: proof of linear conv 2} (at the top of the next page).
%\begin{align}\label{eqn: proof of linear conv 2}
% &\triangle_{k+1} \leq
% \frac{1}{\eta^{k+1}}\triangle_{0} \notag \\
%&+ \frac{1}{ \eta}\bigg[\!\bigg(\!\frac{L^2+(\epsilon-1)\rho+\frac{L^2}{2\rho^2 N}+\frac{2L^2}{\rho}}{2}\bigg)\!\sum_{\ell=0}^k\frac{1}{\eta^{\ell}}\!\sum_{i=1}^N \!\|\xb_i^{k-\ell+1}\!-\!\xb_i^{k-\ell}\|^2
%\notag
%\\
%%\end{align}
%%\begin{align}
%&~~~-
%\bigg(\frac{2\gamma+N\rho}{2}-1\bigg)\sum_{\ell=0}^k\frac{1}{\eta^{\ell}}\|\xb_0^{k-\ell+1}-\xb_0^{k-\ell}\|^2 \notag
%\\
%%\end{align}
%%\begin{align}
%&~~~+  \bigg(\frac{1+\rho/\epsilon}{2}+\frac{1}{2N}\bigg)\underbrace{\sum_{\ell=0}^k\frac{1}{\eta^{\ell}}\sum_{i\in \Ac_{k-\ell}} \|\xb_0^{k-\ell}-\xb_0^{\overline{(k-\ell)}_i+1}\|^2}_{(\ref{eqn: proof of linear conv 2}a)} \notag
%\\
%%\end{align}
%%\begin{align}
%&~~~ +\frac{1}{2N}\underbrace{\sum_{\ell=0}^k\frac{1}{\eta^{\ell}}\sum_{i\in \Ac_{k-\ell}^c}\|\xb_0^{k-\ell}-\xb_0^{\widehat{ (k-\ell)}_i+1}\|^2}_{(\ref{eqn: proof of linear conv 2}b)}\notag \\
%&~~~
%+\frac{ L^2}{4\rho^2 N}\underbrace{\sum_{\ell=0}^k\frac{1}{\eta^{\ell}}\sum_{i\in \Ac_{k-\ell}^c}\|\xb_i^{\widetilde{(k-\ell)}_i+1}-\xb_i^{\widetilde{(k-\ell)}_i}\|_2^2}_{(\ref{eqn: proof of linear conv 2}c)}\bigg].
%\end{align}
%\begin{figure*}
\begin{align}\label{eqn: proof of linear conv 2}
 &\triangle_{k+1} \leq
 \frac{1}{\eta^{k+1}}\triangle_{0} \notag \\
&+ \frac{1}{ \eta}\bigg[\!\bigg(\!\frac{L^2+(\epsilon-1)\rho+\frac{L^2}{2\rho^2 N}+\frac{2L^2}{\rho}}{2}\bigg)\!\sum_{\ell=0}^k\frac{1}{\eta^{\ell}}\!\sum_{i=1}^N \!\|\xb_i^{k-\ell+1}\!-\!\xb_i^{k-\ell}\|^2
-
\bigg(\frac{2\gamma+N\rho}{2}-1\bigg)\sum_{\ell=0}^k\frac{1}{\eta^{\ell}}\|\xb_0^{k-\ell+1}-\xb_0^{k-\ell}\|^2 \notag
\\
%\end{align}
%\begin{align}
&~~~~~~~+  \bigg(\frac{1+\rho/\epsilon}{2}+\frac{1}{2N}\bigg)\underbrace{\sum_{\ell=0}^k\frac{1}{\eta^{\ell}}\sum_{i\in \Ac_{k-\ell}} \|\xb_0^{k-\ell}-\xb_0^{\overline{(k-\ell)}_i+1}\|^2}_{(\ref{eqn: proof of linear conv 2}a)} +\frac{1}{2N}\underbrace{\sum_{\ell=0}^k\frac{1}{\eta^{\ell}}\sum_{i\in \Ac_{k-\ell}^c}\|\xb_0^{k-\ell}-\xb_0^{\widehat{ (k-\ell)}_i+1}\|^2}_{(\ref{eqn: proof of linear conv 2}b)}\notag \\
&~~~~~~~
+\frac{ L^2}{4\rho^2 N}\underbrace{\sum_{\ell=0}^k\frac{1}{\eta^{\ell}}\sum_{i\in \Ac_{k-\ell}^c}\|\xb_i^{\widetilde{(k-\ell)}_i+1}-\xb_i^{\widetilde{(k-\ell)}_i}\|_2^2}_{(\ref{eqn: proof of linear conv 2}c)}\bigg].
\end{align}
%\hrulefill
%\end{figure*}
%where in the third right hand side (RHS) term, we have dropped $-(\frac{2\gamma+\rho}{2}-(2\rho N+1))\|\xb_0^{k+1}-\xb_0^{k}\|^2<0$ as it is negative under \eqref{eqn: linear cond of conv scheme1 2}.

The three terms (\ref{eqn: proof of linear conv 2}a), (\ref{eqn: proof of linear conv 2}b), and (\ref{eqn: proof of linear conv 2}c) in the right hand side (RHS) of \eqref{eqn: proof of linear conv 2} can respectively be bounded as follows, using Lemma \ref{lemma: asyn error bound}. Consider the change of variable $k-\ell=j$. Then,
we have the following chain for (\ref{eqn: proof of linear conv 2}a):
\begin{align}\label{eqn: proof of linear conv 3}
(\ref{eqn: proof of linear conv 2}a)&=\sum_{\ell=0}^k\frac{1}{\eta^{\ell}}\sum_{i\in \Ac_{k-\ell}} \|\xb_0^{k-\ell}-\xb_0^{\overline{(k-\ell)}_i+1}\|^2
\notag \\
&=\frac{1}{\eta^{k}}\sum_{j=0}^k \eta^j\sum_{i\in \Ac_{j}} \|\xb_0^{j}-\xb_0^{\bar{j}_i+1}\|^2 \notag \\
%&=\frac{1}{\eta^{k}}\sum_{j=0}^k \eta^j\sum_{i\in \Ac_{j}} \|\sum_{q=\bar{j}_i+1}^{j-1}(\xb_0^{q}-\xb_0^{q+1})\|^2
%\leq \frac{1}{\eta^{k}}\sum_{j=0}^k \eta^j\sum_{i\in \Ac_{j}} (j-\bar{j}_i-1)\sum_{q=\bar{j}_i+1}^{j-1} \|\xb_0^{q}-\xb_0^{q+1}\|^2 \notag \\
%&\leq  \frac{1}{\eta^{k}}\sum_{j=0}^k \eta^j\sum_{i\in \Ac_{j}} (\tau-1)\sum_{q=j-\tau+1}^{j-1} \|\xb_0^{q}-\xb_0^{q+1}\|^2 \leq  \frac{1}{\eta^{k}}N (\tau-1)\sum_{j=0}^k \eta^j\sum_{q=j-\tau+1}^{j-1} \|\xb_0^{q}-\xb_0^{q+1}\|^2
%\notag \\
&\leq  \frac{1}{\eta^{k}}S (\tau-1)\sum_{j=0}^{k-1} \eta^{j+1}\bigg(\frac{\eta^{\tau-1}-1}{\eta-1}\bigg)\|\xb_0^{j}-\xb_0^{j+1}\|^2
\notag
\\
%\end{align}
%\begin{align}
&= S (\tau-1)\eta\bigg(\frac{\eta^{\tau-1}-1}{\eta-1}\bigg)\sum_{\ell=1}^k \frac{1}{\eta^{\ell}}\|\xb_0^{k-\ell}-\xb_0^{k-\ell+1}\|^2,
\end{align}
where the inequality is obtained by applying \eqref{lemma: asyn error bound 1} with $\nu=\tau$, $\Nc_j=\Ac_j$, $\bar N=S$, and $j_i=\bar j_i$ which satisfies $j-\tau\leq \bar{j}_i <j$ (see \eqref{eqn: cond on bar ki}); to obtain the last equality, the change of variable $k-\ell=j$ is applied again.

Analogously, by applying \eqref{lemma: asyn error bound 1} with $\nu=2\tau-1$, $\Nc_j=\Ac_j^c$, $\bar N=N$, and $j_i=\widehat j_i$ (which satisfies $j-2\tau+1 \leq \widehat j_i <j$ since $\widetilde j_i -\tau \leq \widehat j_i< \widetilde j_i$ and $j-\tau<\widetilde j_i<j$ by \eqref{eqn: cond on bar ki} and \eqref{eqn: cond on bar ki2}),
one can bound (\ref{eqn: proof of linear conv 2}b) as
\begin{align}\label{eqn: proof of linear conv 6}
&(\ref{eqn: proof of linear conv 2}b)=\sum_{\ell=0}^k\frac{1}{\eta^{\ell}}\sum_{i\in \Ac_{k-\ell}^c}\|\xb_0^{k-\ell}-\xb_0^{\widehat{ (k-\ell)}_i+1}\|^2
\notag
\\
%\end{align}
%\begin{align}
&~~~~=\frac{1}{\eta^{k}}\sum_{j=0}^k \eta^j \sum_{i\in \Ac_{j}^c}\|\xb_0^{j}-\xb_0^{\widehat j_i+1}\|^2 \notag
\\
&~~~~ \leq\frac{1}{\eta^{k}}2N(\tau-1)\sum_{j=0}^{k-1} \eta^{j+1} \bigg(\frac{\eta^{2(\tau-1)}-1}{\eta-1}\bigg) \|\xb_0^{j}-\xb_0^{j+1}\|^2 \notag \\
&~~~~= 2N(\tau-1)\eta \bigg(\frac{\eta^{2(\tau-1)}-1}{\eta-1}\bigg) \sum_{\ell=1}^k \frac{1}{\eta^{\ell}}\|\xb_0^{k-\ell}-\xb_0^{k-\ell+1}\|^2.
\end{align}
%where the inequality is obtained by applying \eqref{lemma: asyn error bound 1} with $\nu=2\tau-1$, $\Nc_j=\Ac_j^c$, and $j_i=\hat j_i$ which satisfies $j-2\tau+1 \leq \bar{j}_i <j$ (since $\tilde j_i -\tau \leq \hat j_i< \tilde j_i$ and $j-\tau<\tilde j_i<j$).
The term (\ref{eqn: proof of linear conv 2}c) can be bounded as follows
\begin{align}\label{eqn: proof of linear conv 7}
(\ref{eqn: proof of linear conv 2}c)&=\sum_{\ell=0}^k\frac{1}{\eta^{\ell}}\sum_{i\in \Ac_{k-\ell}^c}\|\xb_i^{\widetilde{(k-\ell)}_i+1}-\xb_i^{\widetilde{(k-\ell)}_i}\|^2
\notag
\\
%\end{align}
%\begin{align}
&=\frac{1}{\eta^{k}}\sum_{j=0}^k \eta^j \sum_{i\in \Ac_{j}^c}\|\xb^{\widetilde j_i+1}_i-\xb_i^{\widetilde j_i}\|^2 \notag \\
&=\frac{1}{\eta^{k}}\sum_{j=0}^k  \sum_{i\in \Ac_{j}^c} \eta^{j-\widetilde j_i-1} \eta^{\widetilde j_i+1} \|\xb^{\widetilde j_i+1}_i-\xb_i^{\widetilde j_i}\|^2
 \notag
% \\
\end{align}
\begin{align}
&\leq\eta^{\tau-2} \frac{1}{\eta^{k}}\sum_{j=0}^k  \sum_{i\in \Ac_{j}^c}  \eta^{\widetilde j_i+1} \|\xb^{\widetilde j_i+1}_i-\xb_i^{\widetilde j_i}\|^2 \notag
\\
%\end{align}
%\begin{align}
&\leq \eta^{\tau-2}(\tau-1)\frac{1}{\eta^{k}}\sum_{i=1}^N \sum_{j=0}^k \eta^{j+1} \|\xb^{j+1}_i-\xb_i^{j}\|^2\notag \\
&=\eta^{\tau-1}(\tau-1)\sum_{i=1}^N \sum_{\ell=0}^k  \frac{1}{\eta^{\ell}}\|\xb^{k-\ell+1}_i-\xb_i^{k-\ell}\|^2,
\end{align}
where, in the first inequality, we have used the fact of $j-\tau+1 \leq \widetilde j_i <j$ from \eqref{eqn: cond on bar ki2}. To show the second inequality,
notice that for any $i\in \Ac_j^c$, it also satisfies $i\in \Ac_\ell^c$ for $\ell=\widetilde j_i+1,\ldots,j$.
So, $\widetilde j_i=\widetilde \ell_i$ for $\ell=\widetilde j_i+1,\ldots,j$.
Since $j-\tau<\widetilde j_i<j$, each $\eta^{\widetilde j+1} \|\xb^{\widetilde j+1}_i-\xb_i^{\widetilde j_i}\|^2$ appears no more than $\tau-1$ times in the summation $\sum_{j=0}^k  \sum_{i\in \Ac_{j}^c}  \eta^{\widetilde j+1} \|\xb^{\widetilde j+1}_i-\xb_i^{\widetilde j_i}\|^2$. %, similar to \eqref{eqn: bound c}.

By substituting \eqref{eqn: proof of linear conv 7}, \eqref{eqn: proof of linear conv 6} and \eqref{eqn: proof of linear conv 3} into \eqref{eqn: proof of linear conv 2}, we obtain
\begin{align}\label{eqn: proof of linear conv 8}
 &\triangle_{k+1} \leq
 \frac{1}{\eta^{k+1}}\triangle_{0}  \notag\\
 &+\frac{1}{ \eta}\bigg[\bigg(\frac{1+\rho/\epsilon}{2}+\frac{1}{2N}\bigg)S (\tau-1)\eta\bigg(\frac{\eta^{\tau-1}-1}{\eta-1}\bigg)
\notag \\
&~~~~~~+(\tau-1)\eta \bigg(\frac{\eta^{2(\tau-1)}-1}{\eta-1}\bigg)\notag
\\
%\end{align}
%\begin{align}
&~~~~~~+1-
\bigg(\frac{2\gamma+N\rho}{2}\bigg)\bigg]\sum_{\ell=0}^k\frac{1}{\eta^{\ell}}\|\xb_0^{k-\ell+1}-\xb_0^{k-\ell}\|^2  \notag
\\
&+\frac{1}{ \eta}\bigg[\bigg(\!\frac{L^2+(\epsilon-1)\rho+\frac{ L^2}{2\rho^2 N}+\frac{2L^2}{\rho}}{2}\bigg) \notag \\
&+\eta^{\tau-1}(\tau-1)\frac{L^2}{4\rho^2 N}\bigg]
\sum_{i=1}^N\sum_{\ell=0}^k\frac{1}{\eta^{\ell}}\sum_{i=1}^N \|\xb_i^{k-\ell+1}-\xb_i^{k-\ell}\|^2.
\end{align}
Let $\epsilon =1/\rho$. Therefore, we see that \eqref{eqn: linear conv rate} is true if
\begin{align}\label{eqn: proof of linear conv 9}
 &\gamma \geq  (\tau-1)\eta\bigg[\bigg(\frac{S(1+\rho^2)+S/N}{2}\bigg)\bigg(\frac{\eta^{\tau-1}-1}{\eta-1}\bigg)
\notag \\
&~~~~~~~~~+\bigg(\frac{\eta^{2(\tau-1)}-1}{\eta-1}\bigg)\bigg]-\frac{N\rho}{2}+1,  \\
 &\rho \geq (1+L^2)+\frac{2L^2}{\rho}+\frac{L^2}{2\rho^2 N}\bigg(1+\eta^{\tau-1}(\tau-1)\bigg).
 \label{eqn: proof of linear conv 10}
\end{align}
Let $\rho\geq \frac{1}{8N}+\sigma^2$. Then \eqref{eqn: proof of linear conv 10} holds true if
\begin{align}
  &\rho \geq (1+L^2)+\frac{2L^2}{\rho}\bigg(1+\frac{2+2\eta^{\tau-1}(\tau-1)}{1+8N\sigma^2}\bigg).
  \label{eqn: proof of linear conv 11}
   %\notag \\
  %\Longleftarrow & \rho\geq \frac{(1+L^2)+\sqrt{(1+L^2)^2+8L^2(1+\frac{4+2\eta^{\tau-1}(\tau-1)}{1+8N\sigma^2})}}{2}
\end{align}
Moreover, since $\gamma\geq 8N(\rho-\sigma^2)$ and $\delta>1$, we see that $\eta$ has an upper bound
\begin{align}\label{eqn: eta}
   \eta =1+\frac{1}{\delta\gamma} < 1+\frac{1}{8N(\rho-\sigma^2)} < 2.
\end{align}
Therefore, \eqref{eqn: linear cond of conv scheme1} and \eqref{eqn: linear cond of conv scheme1 2} are sufficient conditions
for \eqref{eqn: proof of linear conv 11} and \eqref{eqn: proof of linear conv 9}, respectively.
%Therefore, it suffices to have \eqref{eqn: proof of linear conv 9} and \eqref{eqn: proof of linear conv 11} hold true if
%\begin{align}\label{eqn: proof of linear conv 12}
% &\gamma \geq 2N (\tau-1)\bigg[\bigg(\frac{1+\rho^2}{2}+4\rho\bigg)({2^{\tau-1}-1})
%     \notag \\
%     &~~~~~~~~~~~~~~~~~~~+4\rho({4^{\tau-1}-1})\bigg]+\frac{4N-1}{2}\rho+1,  \\
% &\rho \geq \frac{(1+L^2)+\sqrt{(1+L^2)^2+8L^2(2+2^{\tau-1}(\tau-1))}}{2}.
% \label{eqn: proof of linear conv 13}
%\end{align}
%Finally, since \eqref{eqn: proof of linear conv 13} implies $\rho \geq L$, we have
%$\Lc_{\rho}(\xb^{k+1},\xb_0^{k+1}, \lambdab^{k+1}) - F^\star \geq 0$ from Lemma \ref{lemma: Lc lower bounded}.
The proof is thus complete.
\hfill $\blacksquare$

\subsection{Proof of Theorem \ref{thm: linear conv of scheme 1 2}}\label{sec: proof of thm2}

The key is to build a similar result as Lemma \ref{lemma: LC upper bound} under Assumption \ref{assumption scf & osc2}.
Now, consider Assumption \ref{assumption scf & osc2}.
%:
%\begin{align}
%  f_i(\xb_i)=g_i(\Ab_i\xb_i)~i\in \Vc~{\rm and}~h(\xb_0)=0,
%\end{align}
%where $g_i:\mathbb{R}^m\to\mathbb{R}$ is a strongly convex function with a convexity parameter $\sigma^2>0$ and $\Ab_i\in %\mathbb{R}^{m\times n}$ is a nonzero matrix with arbitrary rank, for all $i=1,\ldots,N$.
Let $\xb^\star$ be an optimal solution to \eqref{eqn: original problem}, and let
$$
 \yb_i^\star =\Ab_i\xb^\star,~i=1,\ldots,N.
$$
Then, $(\yb_1^\star,\ldots,\yb_N^\star)$ is unique since $g_i$'s are strongly convex. So, the optimal solution set to \eqref{eqn: consensus problem} can be defined as %{\red[[why? why $h_0$ has nothing to do with the optimality condition?]]}
\begin{align}\label{eqn: opt sol set}
  &\Xc^\star=\bigg\{ (\xb_0,\xb_1,\ldots,\xb_N)~ |~
  \begin{bmatrix}
   \yb_1^\star \\
   \vdots \\
   \yb_N^\star
 \end{bmatrix} =
 \begin{bmatrix}
   \Ab_1\xb_1 \\
   \vdots \\
   \Ab_N\xb_N
 \end{bmatrix}, \notag \\
 &~~~~~~~~~~~~~~~~~~~~~~~~~~~~~~~~\xb_i=\xb_0,~i=1,\ldots,N\bigg\}.
\end{align}
Let $\oneb_{N+1}\otimes \Pc^\star(\hat \xb)$ be the projection point of $\hat\xb\triangleq (\xb_0^T,\xb_1^T,\ldots,\xb_N^T)^T$ onto $\Xc^\star$, where $\otimes$ denotes the Kronecker product.
It can be shown that the following lemma is true.

\begin{Lemma}\label{lemma: osc}
  Under Assumption \ref{assumption scf & osc2}, for any $\hat\xb\in \mathbb{R}^{n(N+1)}$, it holds that
 \begin{align}\label{eqn: OSC 3}
  &\sum_{i=1}^N f_i(\Pc^\star(\hat \xb))
  \geq \sum_{i=1}^N f_i(\xb_i)+ \sum_{i=1}^N (\nabla f_i(\xb_i))^T(\Pc^\star(\hat \xb)-\xb_i)  \notag \\
  &~~~~~~~~~~~~~~~+  \sum_{i=1}^N\frac{\sigma^2}{2c}\|\Pc^\star(\hat \xb)-\xb_i\|^2+\frac{\sigma^2}{2c}\|\Pc^\star(\hat \xb)-\xb_0\|^2 \notag \\
 &~~~~~~~~~~~~~~~- \frac{\sigma^2}{2}\sum_{i=1}^N\|\xb_i-\xb_0\|^2,
\end{align}
for some finite constant $c>0$.
\end{Lemma}
{\bf Proof:} See Appendix \ref{appx: proof of lemma: osc}. \hfill $\blacksquare$

Lemma \ref{lemma: osc} implies that the structured $f_i$'s in Assumption \ref{assumption scf & osc2} own an analogous property as the strongly convex functions in \eqref{eqn: SC}. Based on Lemma \ref{lemma: osc}, the next lemma shows that one can still bound $\triangle_{k+1}$ as in Lemma \ref{lemma: LC upper bound} under Assumption \ref{assumption scf & osc2}.

\begin{Lemma} \label{lemma: LC upper bound 2} Suppose that Assumptions \ref{assumption obj s1} and \ref{assumption scf & osc2} hold, and
assume that $\gamma \geq 8N{(\rho-\sigma^2/c)+4N\sigma^2}$ and $\delta$ satisfies
\begin{align}\label{eqn: liDnear cond of conv scheme1 32}
  %&\frac{ {\rho N}\bigg[(1+\frac{1}{\delta})^2-1\bigg] +\frac{\gamma}{\delta}}{N(1+\frac{1}{\delta})^2} \leq \mu,\\
  &\delta \geq \max\bigg\{1,\frac{\rho N+\gamma}{N\sigma^2/c }-1\bigg\}.
\end{align}
%\begin{align}
%  \frac{ {\rho N}\bigg[(1+\frac{1}{\delta})^2-1\bigg] +\frac{\gamma}{\delta}}{N(1+\frac{1}{\delta})^2} \leq \mu.
%\end{align}
Then, \eqref{eqn: lemma LC upper bound 0} holds true. Instead, if $\gamma=0$ and $\delta\geq \max\{(c\rho)/\sigma^2-1,1\}$, then
\begin{align}\label{eqn: lemma LC upper bound 0 gamma02}
 &\frac{\triangle_{k+1}}{2N[2(\rho-\sigma^2/c) \delta+\sigma^2]} \leq \frac{ L^2}{2\rho^2 N}\sum_{i\in \Ac_k}\|\xb_i^{k+1}-\xb_i^{k}\|^2
  \notag
   \\
  &~~~+ \frac{L^2}{2\rho^2 N}\sum_{i\in \Ac_k^c}\|\xb_i^{\widetilde k_i+1}-\xb_i^{\widetilde k_i}\|^2+\frac{1}{N}\sum_{i\in \Ac_k}\|\xb_0^{k}-\xb_0^{\bar k_i+1}\|^2
  \notag \\
  &~~~+\frac{1}{N}\sum_{i\in \Ac_k^c}\|\xb_0^{k}-\xb_0^{\widehat k_i+1}\|^2 +\|\xb_0^{k+1}-\xb_0^{k}\|^2.
\end{align}

\end{Lemma}
{\bf Proof:} See Appendix \ref{appx: proof of lemma: LC upper bound 2}. \hfill $\blacksquare$
%The proof is presented in Section \ref{appx: proof of lemma: LC upper bound 2}.

Given Lemma \ref{lemma: LC upper bound 2}, Theorem \ref{thm: linear conv of scheme 1 2} can be proved by following exactly the same steps as for Theorem \ref{thm: linear conv of scheme 1} %and Corollary \ref{coro: linear conv of scheme 1}
in Section \ref{subsec: proof of thm1}.
The details are omitted here. \hfill $\blacksquare$

%We end this section with the following remark.

%\subsection{An Alternative Scheme}\label{subsec: alternative scheme}

\section{Numerical Results}
\label{sec: simulations}

\begin{figure*}[!t]
\begin{center}
{\subfigure[][$A=1$]{\resizebox{.47\textwidth}{!}
{\includegraphics{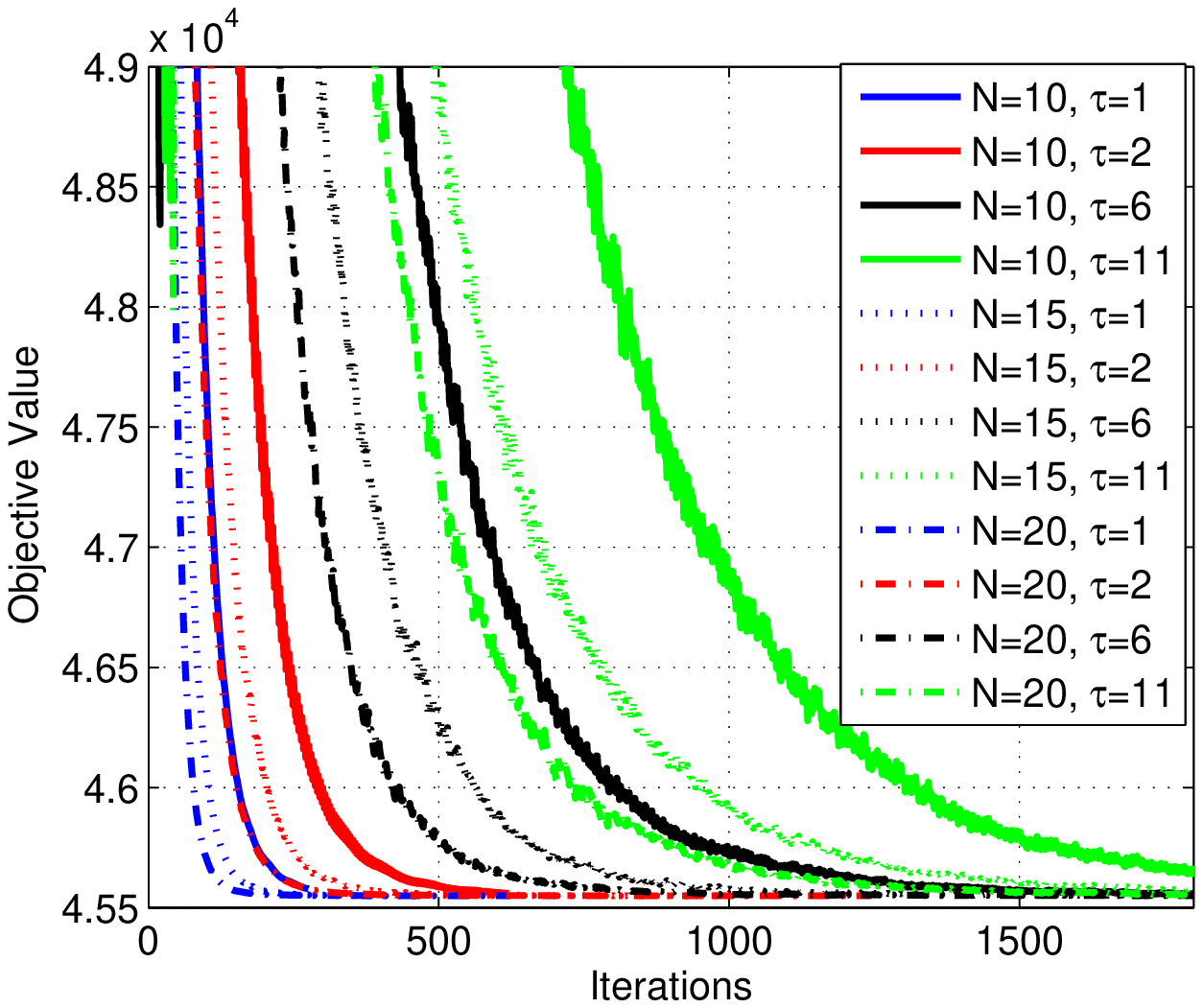}}}
}
\hspace{-1pc}
{\subfigure[][$A=1$]{\resizebox{.47\textwidth}{!}{\includegraphics{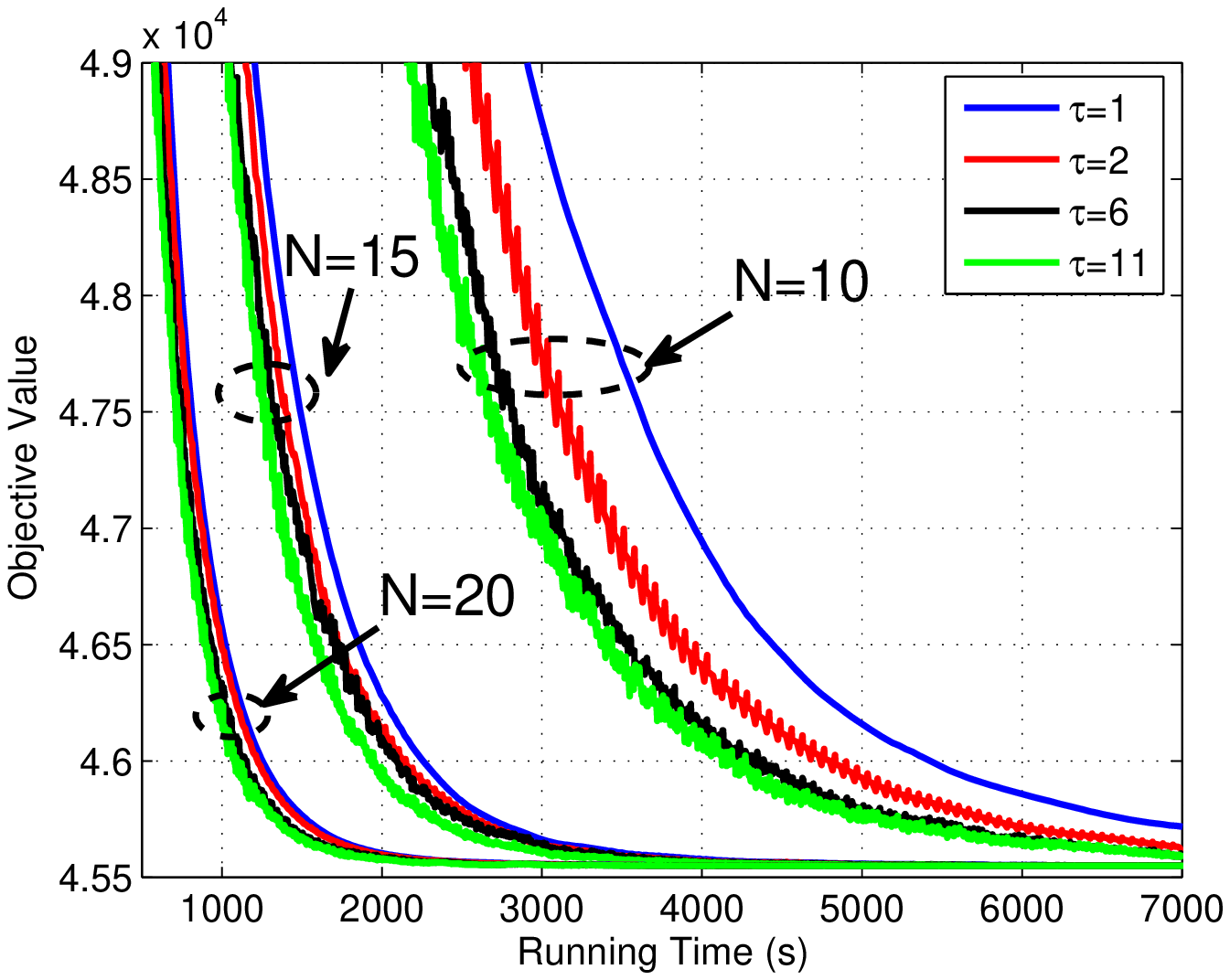}}}
 }
 \hspace{-1pc}
{\subfigure[][$\tau=11$]{\resizebox{.47\textwidth}{!}{\includegraphics{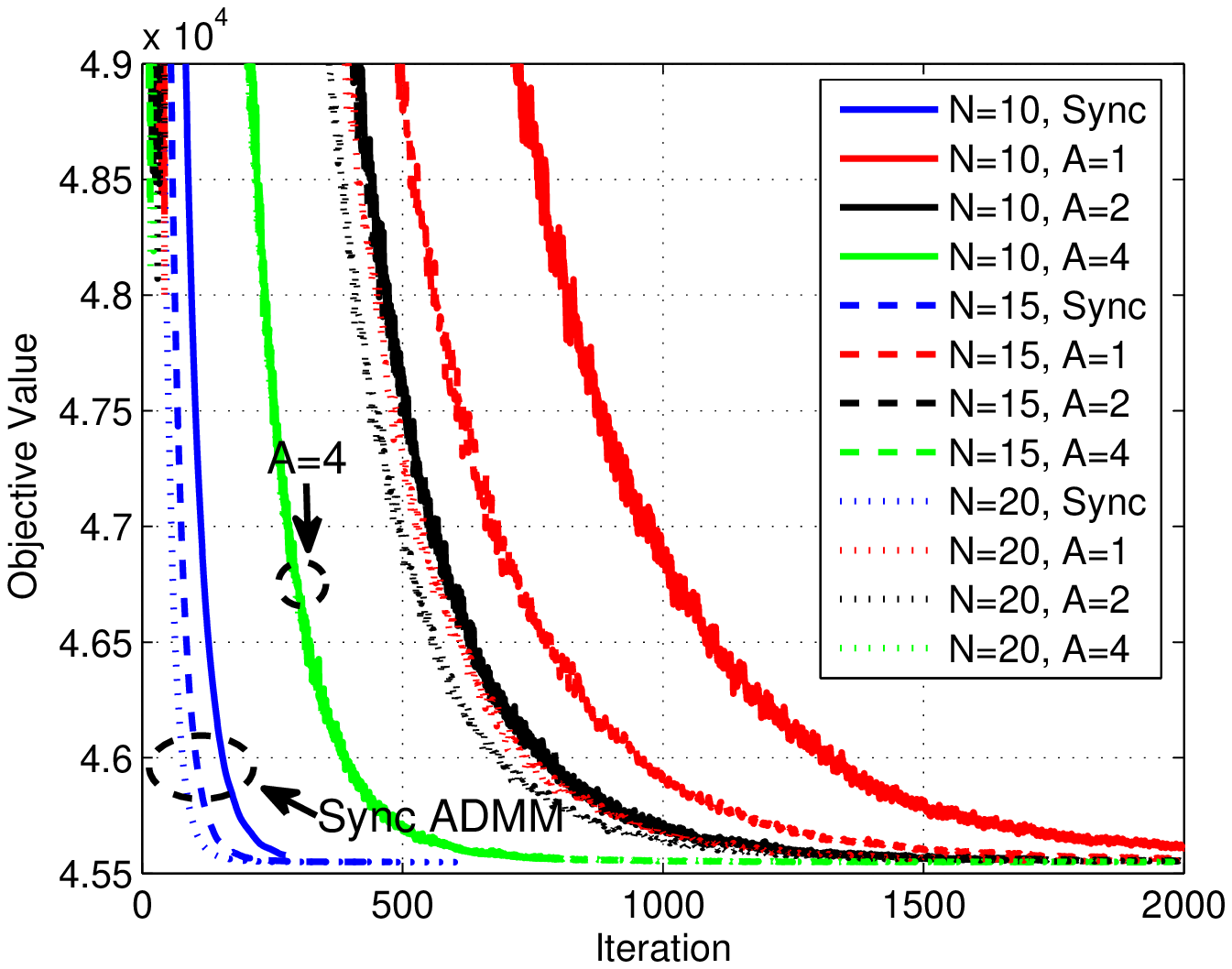}}}
 }
 \hspace{-1pc}
{\subfigure[][$\tau=11$]{\resizebox{.47\textwidth}{!}
{\includegraphics{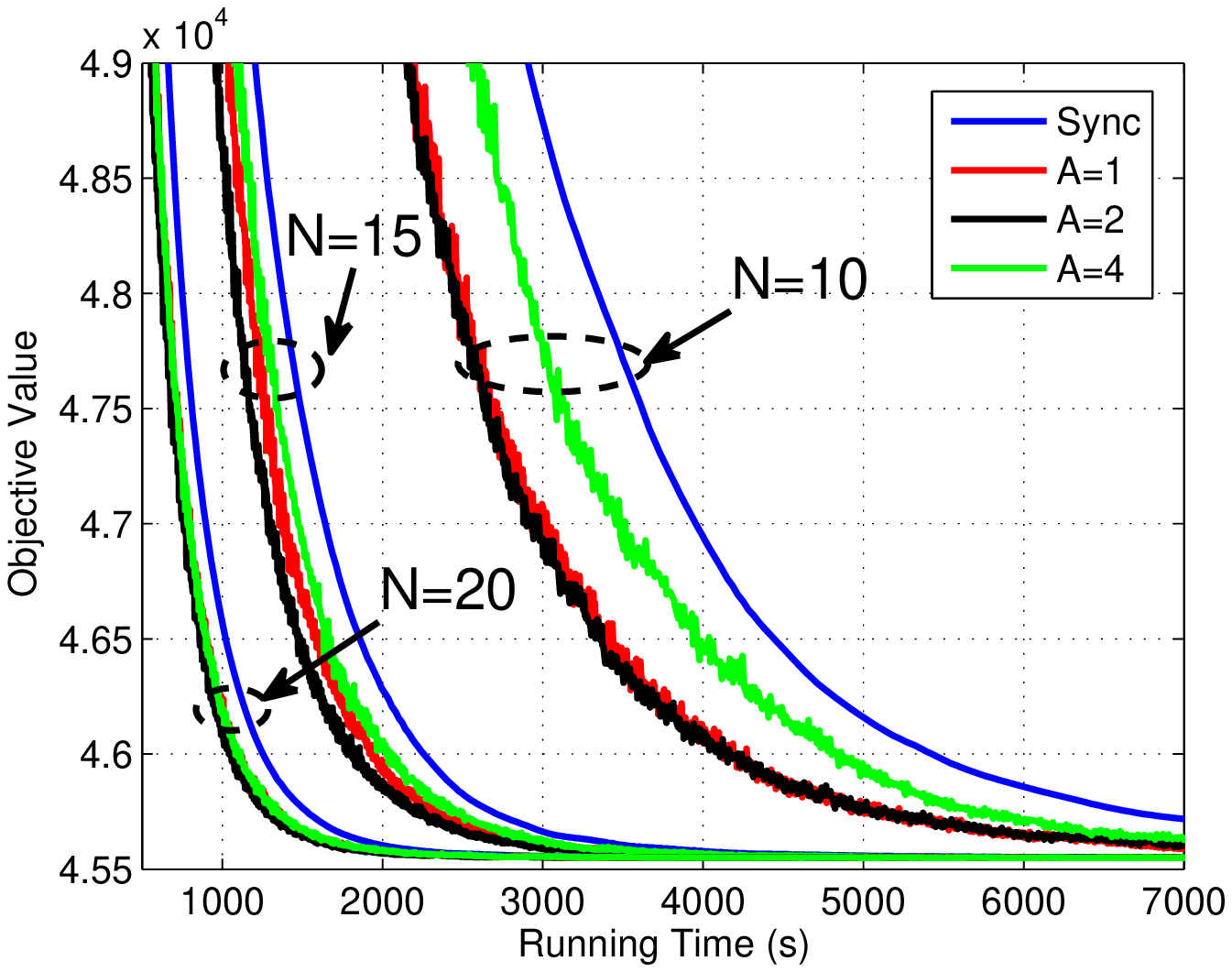}}}
}
\end{center}\vspace{-0.2cm}
\caption{Convergence curves of Algorithm \ref{table: async cadmm s1 master} for solving the LR problem \eqref{LR problem} on the Itasca computer cluster; $\theta=0.1$, $\rho=0.01$ and $\gamma=0$.}
\vspace{-0.2cm}\label{fig: LR conv curves}
\end{figure*}
%\subsection{Example 2: Large data case (Logistic regression)}

In this section, we present some simulation results to examine the practical performance of the AD-ADMM.
We consider the following LR problem
\begin{align}\label{LR problem}
\!\!\!\!\!\min_{\substack{\wb\in \Wc}}&~\sum_{j=1}^{m} \log\big( 1+ \exp(-y_{j}\ab_{j}^T\wb) \big) %\!+\!\theta \|\wb\|_1,
\end{align}
where $y_1,\ldots,y_m$ are the binary labels of the $m$ training data,  $\wb \in \mathbb{R}^{n}$ is the regression variable and $\Ab_i=[\ab_{1},\ldots,\ab_{m}]^T \in \mathbb{R}^{m \times n}$ is the training data matrix.
We used the MiniBooNE particle identification Data Set\footnote{\url{https://archive.ics.uci.edu/ml/datasets/MiniBooNE+particle+identification}}
which contains 130065 training samples ($m=130065$) and the learning parameter has a size of $50$ ($n=50$). The constraint set $\Wc$ is set to $\Wc=\{\wb\in \mathbb{R}^{n}~|~|w_i|\leq 10~ \forall i=1,\ldots,n\}$.
The AD-ADMM is implemented on an HP ProLiant BL280c G6 Linux Cluster (Itasca HPC in University of Minnesota). The $n$ training samples are uniformly distributed to a set of $N$ workers ($N=10,15,20$). For each worker, we employed the fast iterative shrinkage thresholding algorithm (FISTA) \cite{BeckFISTA2009} to solve the corresponding subproblem \eqref{eqn: async cadmm s1 xi}.
The stepsize of FISTA is set to $0.0001$ and the stopping condition is that the 2-norm of the gradient is less than $0.001$. The penalty parameter $\rho$ of the AD-ADMM is set to $0.01$. %and $\gamma=0$.
Interestingly, while the theoretical convergence conditions in \cite[Theorem 1]{ChangAsyncadmm15_p1} and Theorem \ref{thm: linear conv of scheme 1} all suggest that the penalty parameter $\gamma$ should be large in the worst-case, we find that, for the problem instance we test here, it is also fine to set $\gamma=0$.

Note that the asynchrony in our setting comes naturally from the heterogeneity of the computation times of computing nodes. In our experiments, analogous to \cite{Zhang14ACADMM}, we further constrained the minimum size of the active set $\Ac_k$ by $|\Ac_k| \geq A$ where $A\in [1,N]$ is an integer. When $A=N$, it corresponds to the synchronous case where the master is forced to wait for all the workers at every iteration.

\begin{figure*}[!t]
\begin{center}
{\subfigure[][Computation time, $N=10$]{\resizebox{.45\textwidth}{!}
{\includegraphics{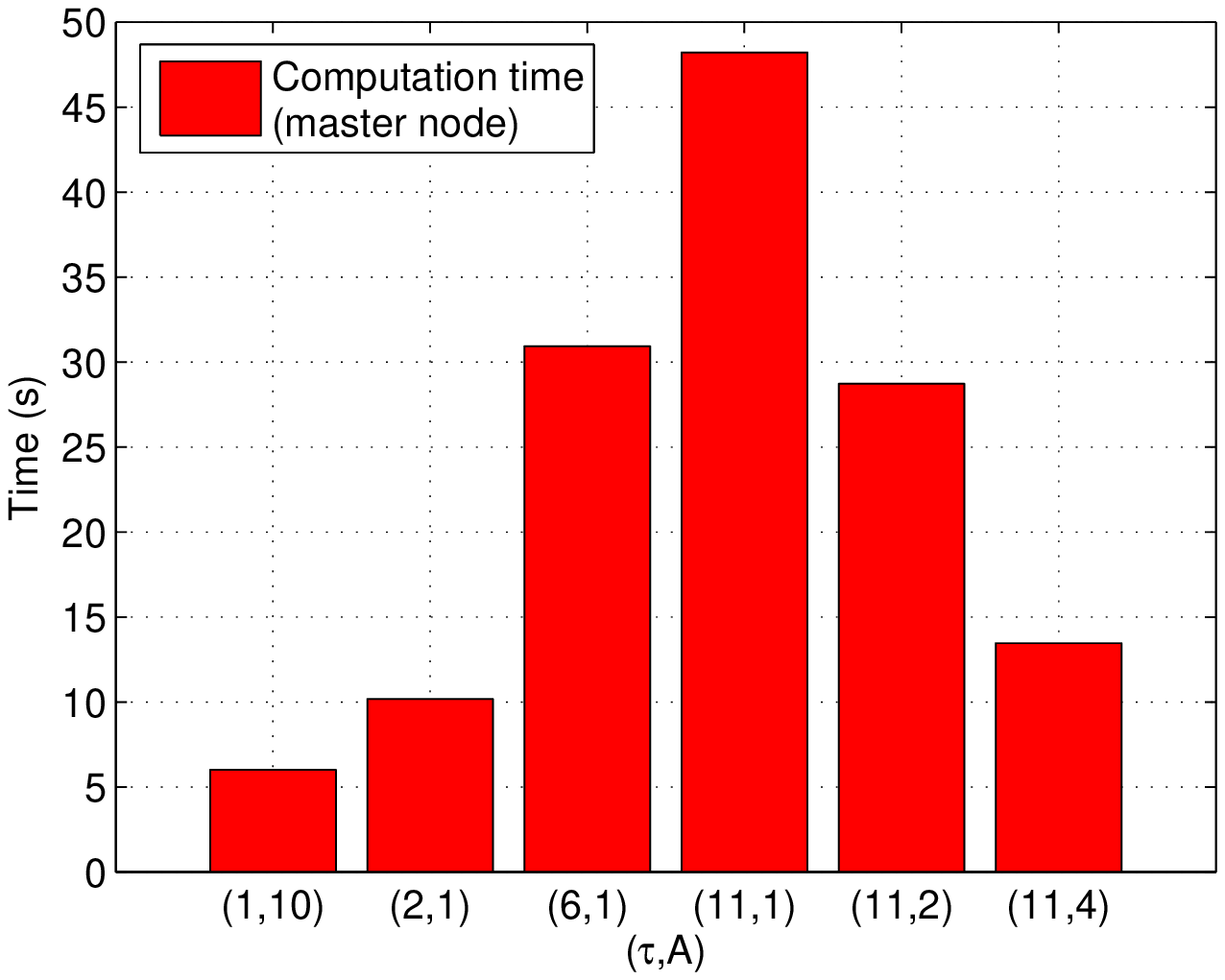}}}
}
\hspace{-1pc}
{\subfigure[][Waiting time, $N=10$]{\resizebox{.45\textwidth}{!}{\includegraphics{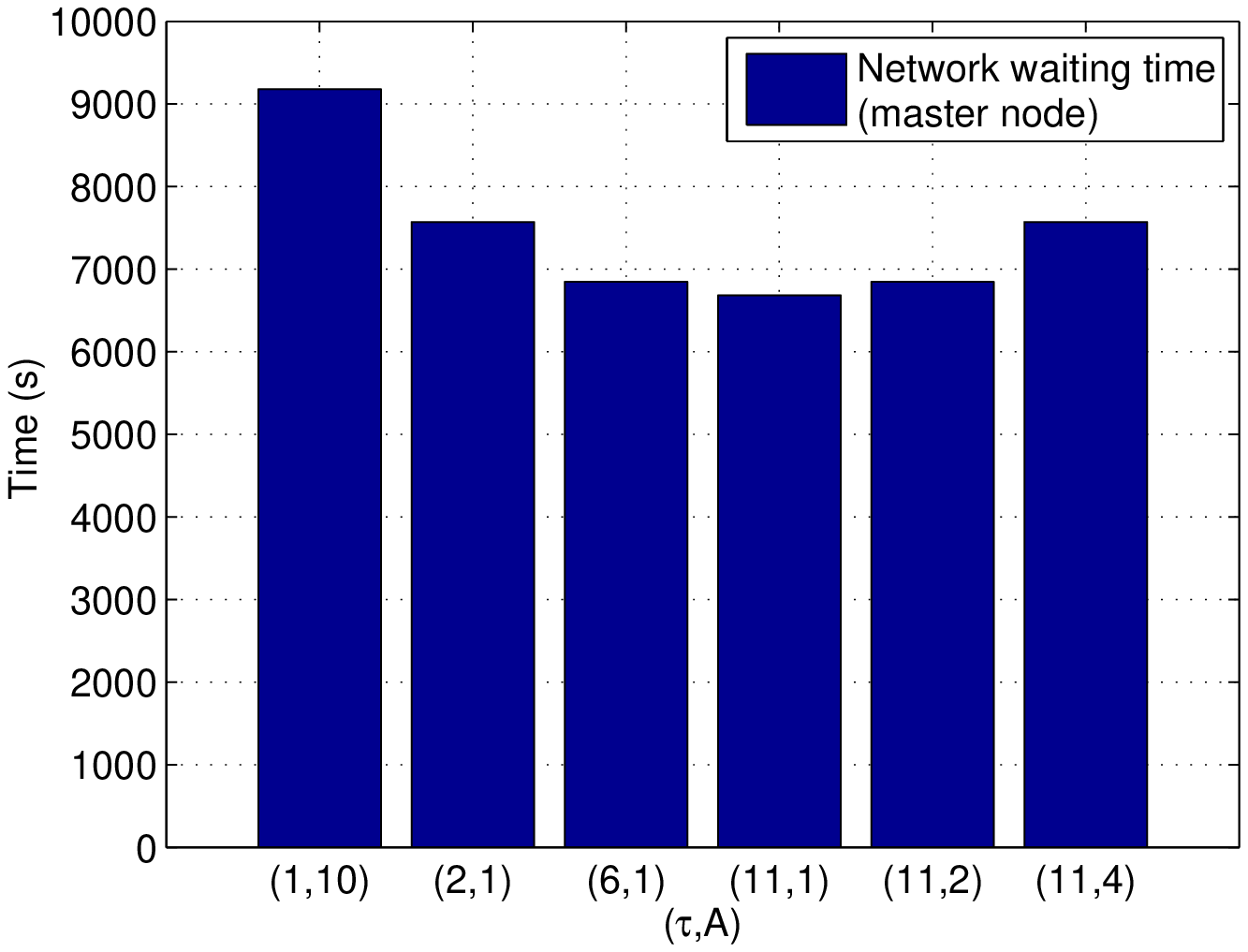}}}
 }
 \hspace{-1pc}
{\subfigure[][Computation time, $N=20$]{\resizebox{.45\textwidth}{!}{\includegraphics{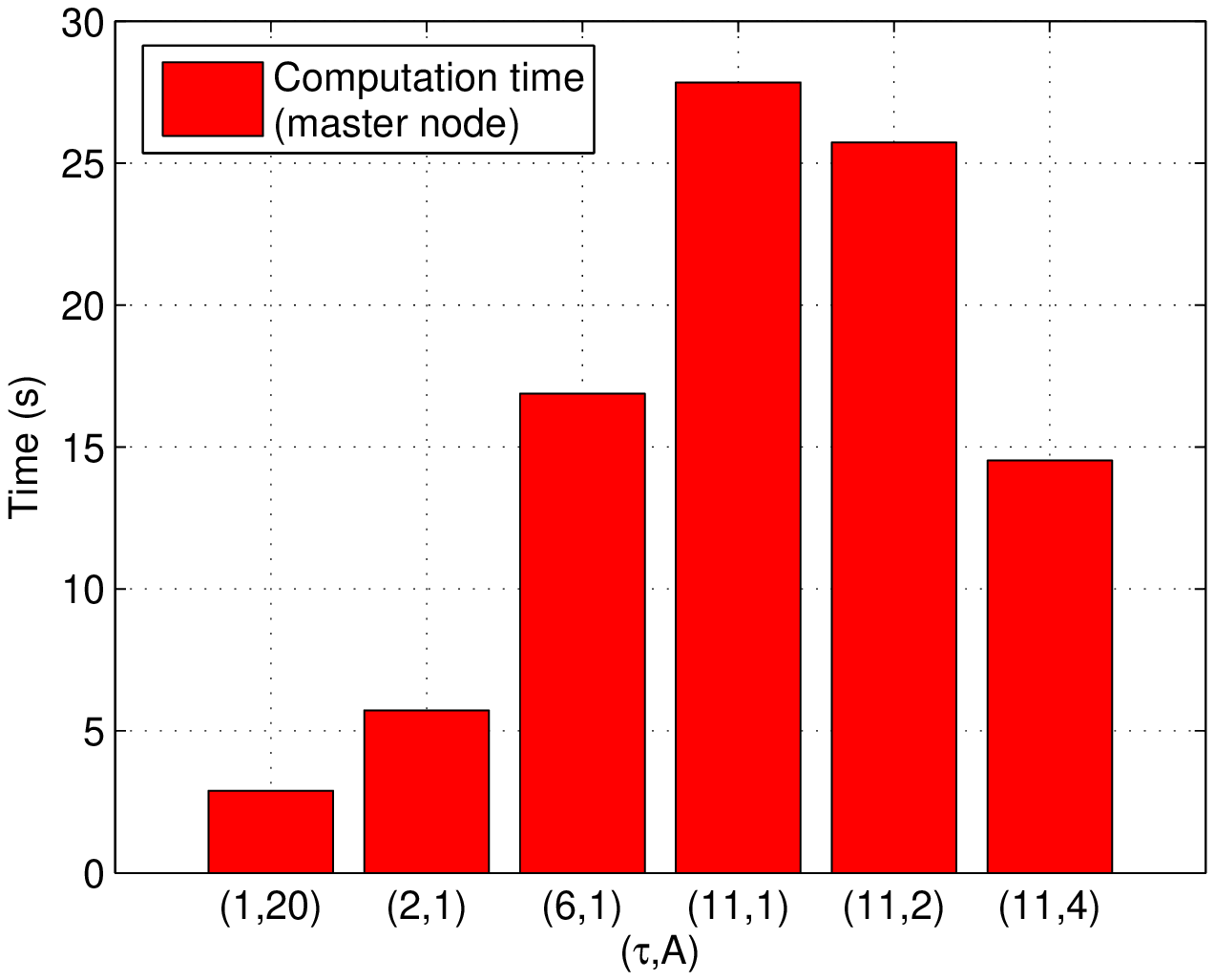}}}
 }
 \hspace{-1pc}
{\subfigure[][Waiting time, $N=20$]{\resizebox{.45\textwidth}{!}
{\includegraphics{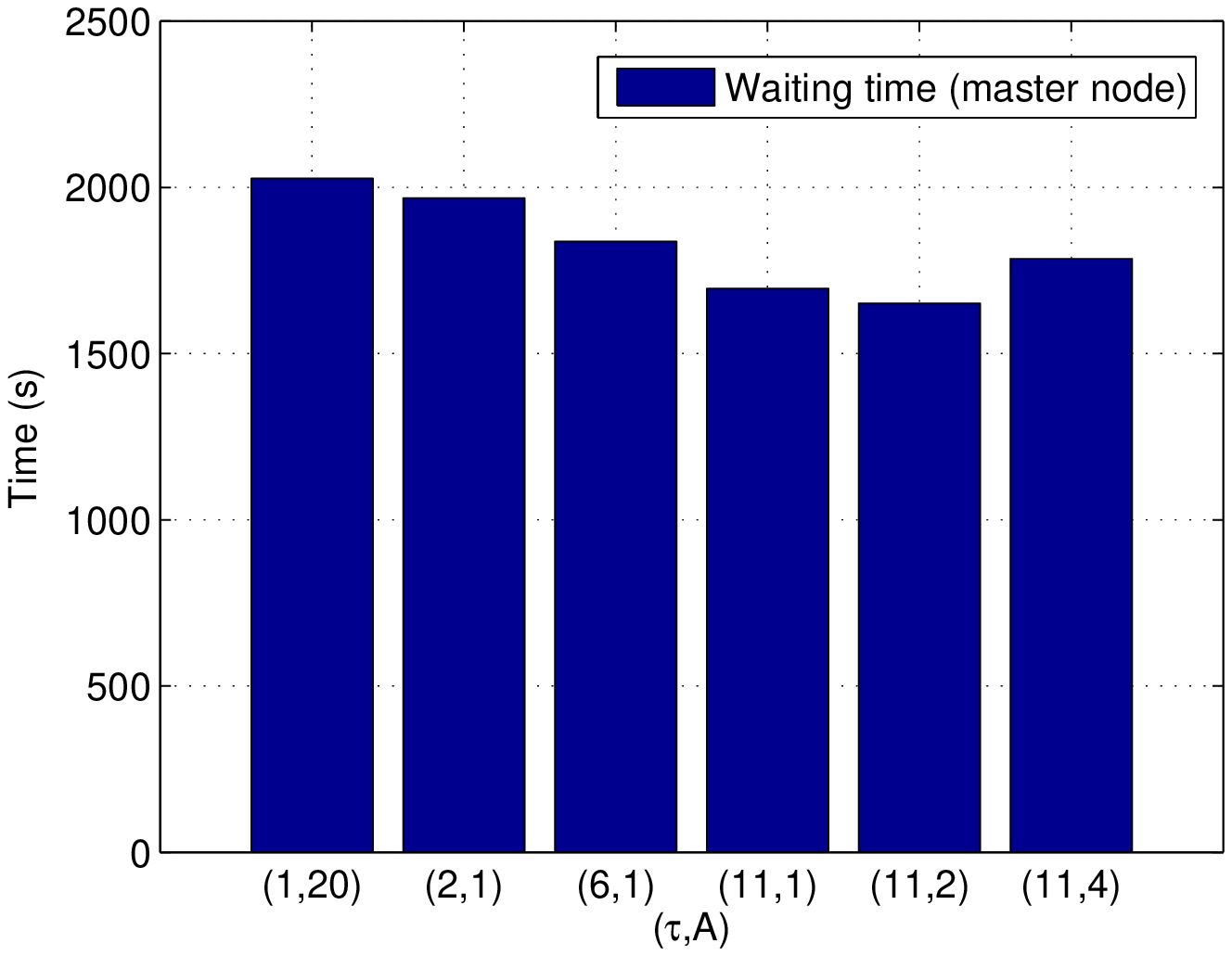}}}
}
\end{center}\vspace{-0.2cm}
\caption{The master's computation and waiting times for solving the LR problem \eqref{LR problem} over the Itasca computer cluster.}
\vspace{-0.0cm}\label{fig: LR time}
\end{figure*}

%\SideNoteMH{Maybe also explain why all ADMM algorithm converges. This connects to our corollary to Theorem 1 (not presented here, but I think we could) that when $\rho$ big enough, the proximal term is not needed. Although in this case we do not know the value of $\tau$, but this may be useful.}

Figure \ref{fig: LR conv curves}(a) and Figure \ref{fig: LR conv curves}(b) respectively display the convergence curves (objective value) of the AD-ADMM versus the iteration number and the running time (second), for various values of $N$ and $\tau$. Here we set $A=1$. One can observe from Figure \ref{fig: LR conv curves}(a) that, in terms of the iteration number, the convergence speed of the AD-ADMM slows down when $\tau$ increases. However, as seen from Figure \ref{fig: LR conv curves}(b), the AD-ADMM is actually faster than its synchronous counterpart ($\tau=1$), and the running time of the AD-ADMM can be further reduced with increased $\tau$. We also observe that, when $N$ increases, the advantages of the AD-ADMM compared to its synchronous counterpart reduces. This is because the computation load allocated to each worker decreases with $N$ (as $n$ is fixed), making all the workers experience similar computation delays. % in solving the local subproblems.

In Figure \ref{fig: LR conv curves}(c) and Figure \ref{fig: LR conv curves}(d), we present the convergence curves of AD-ADMM with different values of $A$. We see from Figure \ref{fig: LR conv curves}(c) that when $A$ increases, it always requires fewer number of iterations to achieve convergence for all choices of parameters. From Figure \ref{fig: LR conv curves}(d), however we can observe that a larger value of $A$ is not always beneficial in reducing the running time. Specifically, one can see that for $N=10$, the running time of AD-ADMM decreases when one increases $A$ from $1$ to $2$, whereas the running time increases a lot if one increases $A$ to $4$. One can observe similar results for $N=15$ and $N=20$.

To look into how the values of $\tau$ and $A$ impact on the algorithm speed, in Figure \ref{fig: LR time}, we respectively plot the computation time and the waiting time of the master node for various pairs of $(\tau,A)$. %In particular, Figure \ref{fig: LR time}(a) and Figure \ref{fig: LR time}(b) show the results for $N=10$ and Figure \ref{fig: LR time}(c) and Figure \ref{fig: LR time}(d) show the results for $N=20$.
The setting is the same as that in Figure \ref{fig: LR conv curves}, except that here the stopping condition of the AD-ADMM is that the objective value achieves $4.56 \times 10^4$. %{\red[[where does this number come from; also what's the stopping criteria for the previous case?]]}.
One can observe from these figures that, when $\tau$ increases, the computing load of the master also increases but the waiting time is significantly reduced. This explains why in Figure \ref{fig: LR conv curves}(b) the AD-ADMM requires a less running time compared with the synchronous ADMM.
On the other hand, when $A$ increases, the computation time of the master always decreases. This is because the master may take a smaller number of iterations to reach the target objective value (see Figure \ref{fig: LR conv curves}(c)) and have to spend more time waiting for slow workers. However, the overall waiting time of the master does not necessarily become larger or smaller with $A$.
As seen from Figure \ref{fig: LR time}(b) and Figure \ref{fig: LR time}(d), when $A$ increases from 1 to 2, the waiting time for $N=10$ in Figure \ref{fig: LR time}(b) increases, whereas the waiting time for $N=20$ in Figure \ref{fig: LR time}(d) decreases. However, for $A=4$, the waiting times always become larger. Nevertheless, when comparing to the synchronous ADMM (i.e., $(\tau,A)=(1,N)$), we can see that the waiting time of the master in the AD-ADMM is always much smaller.

\section{Conclusions}\label{sec: conclusions}
In this paper, we have analytically studied the linear convergence conditions of the AD-ADMM proposed in \cite{ChangAsyncadmm15_p1}.
Specifically, we have shown that for strongly convex $f_i$'s (Assumption \ref{assumption scf & osc}) or for $f_i$'s with the composite form in Assumption \ref{assumption scf & osc2}, the AD-ADMM is guaranteed to converge linearly, provided that the penalty parameter $\rho$ and the proximal parameter $\gamma$ are chosen sufficiently large depending on the delay $\tau$. When the delay $\tau$ is bounded and $N$ is large, we have further shown that linear convergence can be achieved with zero proximal parameter (i.e.,$\gamma=0$), and with a delay-independent $\rho$.
The linear convergence conditions and the linear rate have been given explicitly, which relate the algorithm and network parameters with the algorithm worst-case convergence performance.
The presented numerical examples have shown that in practice the AD-ADMM can effectively reduce the waiting time of the master node, and as a consequence improves the overall time efficiency of distributed optimization significantly.

%As a worst-case analysis, the presented convergence conditions are somehow conservative,

\vspace{-0.0cm}
\appendices {\setcounter{equation}{0}
\renewcommand{\theequation}{A.\arabic{equation}}

\section{Bound of Consensus Error}\label{appx: proof of lemma: consensus error}
%{\bf Proof of Lemma \ref{lemma: consensus error}:}

We bound the size of the consensus error $\sum_{i=1}^N\|\xb_i^{k+1}\!-\!\xb_0^{k+1}\|^2$ %by successive differences of variables $\xb_i^k$, $i\in \Vc$, and $\xb_0^k$
in the following lemma.

\begin{Lemma} \label{lemma: consensus error} Under Assumption \ref{assumption obj s1}, it holds that
\begin{align}\label{eqn: consensus error bound}
\!\!\!\!\!  &\sum_{i=1}^N\|\xb_i^{k+1}\!-\!\xb_0^{k+1}\|^2\leq \frac{2L^2}{\rho^2}\sum_{i\in \Ac_k}\|\xb_i^{k+1}\!-\xb_i^{k}\|^2
\notag
\\
%\end{align}
%\begin{align}
\!\!\!\!\!  &
   ~~~+\frac{2L^2}{\rho^2}\sum_{i\in \Ac_k^c}\|\xb_i^{\widetilde k_i+1}\!-\xb_i^{\widetilde k_i}\|^2
   +4\sum_{i\in \Ac_k}\|\xb_0^{\bar k_i+1}-\!\xb_0^{k}\|^2+4\sum_{i\in \Ac_k^c}\|\xb_0^{\widehat k_i+1}-\!\xb_0^{k}\|^2+4N\|\xb^{k+1}_0-\xb_0^k\|^2.
\end{align}
\end{Lemma}

%The proof is presented in Section \ref{appx: proof of lemma: consensus error}.
{\bf Proof:} It follows from \eqref{eqn: async cadmm s1 lambda equi2} ~and ~\eqref{eqn: async cadmm s1 lambda skc equi} that the following chain is true
\begin{align}\label{eq: proof 20}
   &\sum_{i=1}^N\|\xb_i^{k+1}\!-\!\xb_0^{k+1}\|^2=\!
   \sum_{i\in \Ac_k}\!\!\|\xb_i^{k+1}\!-\xb_0^{\bar k_i+1}+\xb_0^{\bar k_i+1}-\!\xb_0^{k+1}\|^2
   \notag
   \\
   %\end{align}
   %\begin{align}
   &~~~~~~~~~~~~+\sum_{i\in \Ac_k^c}\|\xb_i^{k+1}\!-\xb_0^{\widehat k_i+1}+\xb_0^{\widehat k_i+1}-\!\xb_0^{k+1}\|^2
   \notag
   \\
   %\end{align}
   %\begin{align}
   & \leq
   2\sum_{i\in \Ac_k}\|\xb_i^{k+1}\!-\xb_0^{\bar k_i+1}\|^2+2\sum_{i\in \Ac_k^c}\|\xb_i^{k+1}\!-\xb_0^{\widehat k_i+1}\|^2 \notag \\
   &~~~~~+2\sum_{i\in \Ac_k}\|\xb_0^{\bar k_i+1}-\!\xb_0^{k+1}\|^2+2\sum_{i\in \Ac_k^c}\|\xb_0^{\widehat k_i+1}-\!\xb_0^{k+1}\|^2 \notag
   \\
   %\end{align}
   %\begin{align}
   &\leq
      \frac{2}{\rho^2}\sum_{i\in \Ac_k}\|\lambdab_i^{k+1}\!-\lambdab_i^{k}\|^2
   +\frac{2}{\rho^2}\sum_{i\in \Ac_k^c}\|\lambdab_i^{\widetilde k_i+1}\!-\lambdab_i^{\widetilde k_i}\|^2 \notag \\
   &~~~+2\sum_{i\in \Ac_k}\|\xb_0^{\bar k_i+1}-\xb_0^k+\xb_0^k-\!\xb_0^{k+1}\|^2 +2\sum_{i\in \Ac_k^c}\|\xb_0^{\widehat k_i+1}-\xb_0^k+\xb_0^k-\!\xb_0^{k+1}\|^2
   \notag
   \\
   %\end{align}
   %\begin{align}
   &\leq
      \frac{2}{\rho^2}\sum_{i\in \Ac_k}\|\lambdab_i^{k+1}\!-\lambdab_i^{k}\|^2
   +\frac{2}{\rho^2}\sum_{i\in \Ac_k^c}\|\lambdab_i^{\widetilde k_i+1}\!-\lambdab_i^{\widetilde k_i}\|^2 \notag \\
   &~~+4\sum_{i\in \Ac_k}\|\xb_0^{\bar k_i+1}-\!\xb_0^{k}\|^2+4\sum_{i\in \Ac_k^c}\|\xb_0^{\widehat k_i+1}-\!\xb_0^{k}\|^2
   +4N\|\xb^{k+1}_0-\xb_0^k\|^2.
   \end{align}
Recall from \cite[Eqn. (38)]{ChangAsyncadmm15_p1} that
\begin{align}\label{lemma: Lc progress eq 8}
 \nabla f_i(\xb_i^{k+1})+\lambdab_i^{k+1}=\zerob~\forall i\in \Vc ~{\rm and}~ \forall k.
\end{align}
By substituting \eqref{lemma: Lc progress eq 8} into \eqref{eq: proof 20} and by the Lipschitz continuity of $\nabla f_i$, we obtain \eqref{eqn: consensus error bound}.
\hfill $\blacksquare$

\section{Proof of Lemma \ref{lemma: asyn error bound}}\label{appx: proof of lemma: asyn error bound}

%{\bf Proof of Lemma \ref{lemma: asyn error bound}:}
It is easy to show the following chain is true
\begin{align}\label{eqn: proof of linear conv 30}
&\sum_{j=0}^k \eta^j\!\sum_{i\in \Nc_{j}}\!\! \|\xb_0^{j}-\xb_0^{j_i+1}\|^2  =\!\sum_{j=0}^k \eta^j\sum_{i\in \Nc_{j}} \|\!\sum_{q=j_i+1}^{j-1}\!(\xb_0^{q}-\xb_0^{q+1})\|^2 \notag \\
&~~\leq \sum_{j=0}^k \eta^j\sum_{i\in \Nc_{j}} (j-j_i-1)\sum_{q=j_i+1}^{j-1} \|\xb_0^{q}-\xb_0^{q+1}\|^2 \notag \\
&~~\leq  \sum_{j=0}^k \eta^j\sum_{i\in \Nc_{j}} (\nu-1)\sum_{q=j-\nu+1}^{j-1} \|\xb_0^{q}-\xb_0^{q+1}\|^2 \notag \\
&~~ \leq   (\nu-1)\bar N\bigg(\sum_{j=0}^k \eta^j\sum_{q=j-\nu+1}^{j-1} \|\xb_0^{q}-\xb_0^{q+1}\|^2\bigg),
%\notag \\
%&\leq  \frac{1}{\eta^{k}}N (\nu-1)\sum_{j=0}^k \eta^{j+1}\bigg(\frac{\eta^{\nu-1}-1}{\eta-1}\bigg)\|\xb_0^{j}-\xb_0^{j+1}\|^2
%\notag \\
%&= N (\nu-1)\eta\bigg(\frac{\eta^{\nu-1}-1}{\eta-1}\bigg)\sum_{\ell=0}^k \frac{1}{\eta^{\ell}}\|\xb_0^{k-\ell}-\xb_0^{k-\ell+1}\|^2,
\end{align}
where the second inequality is owing to $j-\nu\leq j_i$. To proceed, we list $\eta^j \sum_{q=j-\nu+1}^{j-1} \|\xb_0^{q}-\xb_0^{q+1}\|^2$ for $j=1,\ldots,\nu,\ldots$, below %in \eqref{eqn: proof of linear conv 50} at the top of the next page.
%\begin{figure*}
\begin{align}\label{eqn: proof of linear conv 50}
  %&j=0,    ~~~~~~~~~~~~~~  ~~ 0 \notag \\
  &j=1,    ~~~~~  ~~ \eta \sum_{q=2-\nu}^{0} \|\xb_0^{q}-\xb_0^{q+1}\|^2 = \eta \|\xb_0^{0}-\xb_0^{1}\|^2 \notag \\
  &j=2,    ~~~~~  ~~   \eta^2 \sum_{q=3-\nu}^{1} \|\xb_0^{q}-\xb_0^{q+1}\|^2 =  \eta^2 \|\xb_0^{0}-\xb_0^{1}\|^2 +\eta^2 \|\xb_0^{1}-\xb_0^{2}\|^2 \notag \\
  &~~~~\vdots ~~~~~~~~~~~~~~~~~~~~~~~ \vdots \notag\\
  &j=\nu-1,    ~~   \eta^{\nu-1} \sum_{q=0}^{\nu-2} \|\xb_0^{q}-\xb_0^{q+1}\|^2= \eta^{\nu-1} \|\xb_0^{0}-\xb_0^{1}\|^2 +\eta^{\nu-1} \|\xb_0^{1}-\xb_0^{2}\|^2
                                +\cdots +\eta^{\nu-1} \|\xb_0^{\nu-2}-\xb_0^{\nu-1}\|^2 \notag \\
  &j=\nu,    ~~~~~~~  \eta^\nu \sum_{q=1}^{\nu-1} \|\xb_0^{q}-\xb_0^{q+1}\|^2=  \eta^{\nu} \|\xb_0^{1}-\xb_0^{2}\|^2
   +\eta^{\nu} \|\xb_0^{2}-\xb_0^{3}\|^2
                                +\cdots +\eta^{\nu} \|\xb_0^{\nu-1}-\xb_0^{\nu}\|^2  %\notag %\\
  %&~~~~\vdots ~~~~~~~~~~~~~~~~~~~~~~~ \vdots
\end{align}
%\hrulefill
%\end{figure*}
One can verify that each $\|\xb_0^{j}-\xb_0^{j+1}\|^2$ appears no more than $\nu-1$ times in the summation term $\sum_{j=0}^k \eta^j\sum_{q=j-\nu+1}^{j-1} \|\xb_0^{q}-\xb_0^{q+1}\|^2$ and therefore the total contribution of each $\|\xb_0^{j}-\xb_0^{j+1}\|^2$ can be upper bounded by
\begin{align}
  &(\eta^{j+1} + \eta^{j+2} +\cdots+\eta^{j+\nu-1}) \|\xb_0^{j}-\xb_0^{j+1}\|^2
  =\eta^{j+1}\bigg(\frac{\eta^{\nu-1}-1}{\eta-1}\bigg)\|\xb_0^{j}-\xb_0^{j+1}\|^2.
\end{align} This shows that
\begin{align}\label{eqn: proof of linear conv 4}
&\sum_{j=0}^k \eta^j\sum_{q=j-\nu+1}^{j-1} \|\xb_0^{q}-\xb_0^{q+1}\|^2 \leq \sum_{j=0}^{k-1} \eta^{j+1}\bigg(\frac{\eta^{\nu-1}-1}{\eta-1}\bigg)\|\xb_0^{j}-\xb_0^{j+1}\|^2.
\end{align}
By substituting \eqref{eqn: proof of linear conv 4} into \eqref{eqn: proof of linear conv 30}, we obtain \eqref{lemma: asyn error bound 1}. %The inequality \eqref{lemma: asyn error bound 1} can be proved in the same way.
\hfill $\blacksquare$

\section{Proof of Lemma \ref{lemma: LC upper bound}}\label{appx: proof of lemma: LC upper bound}
%{\bf Proof of Lemma \ref{lemma: LC upper bound} :}
By the optimality condition of \eqref{eqn: async cadmm s1 xi equi2} \cite{BK:BoydV04} , one has, $\forall i\in \Ac_k$ and $\forall \xb_i\in \mathbb{R}^n$,
\begin{align}\label{eqn: proof of lemma Lc upper bound 1}
\!\!\!\!\!  0&\geq   (\nabla f_i(\xb_i^{k+1})+\lambdab^k_i +\rho(\xb_i^{k+1} -\xb_0^{\bar k_i +1})^T(\xb_i^{k+1}-\xb_i) \notag \\
   &= (\nabla f_i(\xb_i^{k+1}))^T(\xb_i^{k+1}-\xb_i)+(\lambdab^{k +1}_i)^T(\xb_i^{k+1}-\xb_i),
\end{align}where the equality is due to \eqref{eqn: async cadmm s1 lambda equi2}.
Similarly, by the optimality condition of \eqref{eqn: async cadmm s1 xi skc equi} and by \eqref{eqn: async cadmm s1 lambda skc equi}, one has, $\forall i\in \Ac_k^c$ and $\forall \xb_i\in \mathbb{R}^n$,
\begin{align}\label{eqn: proof of lemma Lc upper bound 2}
  0&\geq   (\nabla f_i(\xb_i^{k+1})+\lambdab^{\widetilde k_i }_i +\rho(\xb_i^{k+1} -\xb_0^{\widehat k_i +1})^T(\xb_i^{k+1}-\xb_i) \notag \\
   &= (\nabla f_i(\xb_i^{k+1}))^T(\xb_i^{k+1}-\xb_i)+(\lambdab^{k +1}_i)^T(\xb_i^{k+1}-\xb_i).
\end{align}
Summing \eqref{eqn: proof of lemma Lc upper bound 1} and \eqref{eqn: proof of lemma Lc upper bound 2} for all $i\in \Vc$ gives rise to
\begin{align}\label{eqn: proof of lemma Lc upper bound 3}
   & \sum_{i=1}^N(\nabla f_i(\xb_i^{k+1}))^T(\xb_i^{k+1}-\xb_i) +\sum_{i=1}^N(\lambdab^{k +1}_i)^T(\xb_i^{k+1}-\xb_i)\notag \\
   &\leq 0~~\forall (\xb_1,\ldots,\xb_N)\in \mathbb{R}^{nN}.
\end{align}
In addition, by the optimality condition of \eqref{eqn: async cadmm s1 x0 equi2} \cite[Lemma 4.1]{BertsekasADMM}, one has, $\forall \xb_0\in \mathbb{R}^n$,
\begin{align}\label{eqn: proof of lemma Lc upper bound 4}
& h(\xb_0^{k+1})-  h(\xb_0)- \sum_{i=1}^N (\lambdab_i^{k+1})^T(\xb_0^{k+1}-\xb_0)
\notag \\
&~~~~~~~~~~ -{\rho}\sum_{i=1}^N(\xb_i^{k+1}-\xb_0^{k+1})^T(\xb_0^{k+1}-\xb_0) \notag \\
&~~~~~~~~~~+{\gamma}(\xb_0^{k+1}-\xb_0^k)^T(\xb_0^{k+1}-\xb_0)\leq 0.
\end{align}
Denote $\xb^\star\in \mathbb{R}^n$ as an optimal solution to problem \eqref{eqn: original problem}. Let
$\xb_1=\cdots=\xb_N=\xb_0=\xb^\star$ in \eqref{eqn: proof of lemma Lc upper bound 3} and \eqref{eqn: proof of lemma Lc upper bound 4}, and combine the two equations. We obtain
\begin{align}
 &\sum_{i=1}^N(\nabla f_i(\xb_i^{k+1}))^T(\xb_i^{k+1}-\xb^\star ) +
  h(\xb_0^{k+1})-  h(\xb^\star) \notag \\
  &~~~+\sum_{i=1}^N(\lambdab^{k +1}_i)^T(\xb_i^{k+1}-\xb_0^{k+1})
 \notag \\
 &~~~ -{\rho}\sum_{i=1}^N(\xb_i^{k+1}-\xb_0^{k+1})^T(\xb_0^{k+1}-\xb^\star ) \notag \\
 &~~~+{\gamma}(\xb_0^{k+1}-\xb_0^k)^T(\xb_0^{k+1}-\xb^\star )\leq 0. \label{eqn: proof of lemma Lc upper bound 510}
 %\\
 \end{align}

Let $\yb=\xb^\star$ and $\xb=\xb_i^{k+1}$ in \eqref{eqn: SC} for all $i\in \Vc$, and apply them to \eqref{eqn: proof of lemma Lc upper bound 510}. We have
\begin{align}
 0&\geq \bigg(\sum_{i=1}^N f_i(\xb_i^{k+1})+h(\xb_0^{k+1}) - \sum_{i=1}^N f_i(\xb^\star)
 -h(\xb^\star)\bigg) \notag \\
 &~~~+\sum_{i=1}^N(\lambdab^{k +1}_i)^T(\xb_i^{k+1}-\xb_0^{k+1})+\frac{\sigma^2}{2}\sum_{i=1}^N\|\xb_i^{k+1}-\xb^\star\|^2
 \notag \\
 &~~~
 -{\rho}\sum_{i=1}^N(\xb_i^{k+1}-\xb_0^{k+1})^T(\xb_0^{k+1}-\xb^\star
 )\notag\\
 &~~~+{\gamma}(\xb_0^{k+1}-\xb_0^k)^T(\xb_0^{k+1}-\xb^\star ),\label{eqn: proof of lemma Lc upper bound 512}
\end{align}
Note that, by \eqref{eqn: identity},
\begin{align}\label{eqn: proof of lemma Lc upper bound 6}
  &-{\rho}\sum_{i=1}^N(\xb_i^{k+1}-\xb_0^{k+1})^T(\xb_0^{k+1}-\xb^\star)\notag \\
  &
  =-\frac{\rho}{2}\sum_{i=1}^N\|\xb_i^{k+1}-\xb^\star\|^2+\frac{\rho}{2}\sum_{i=1}^N\|\xb_i^{k+1}-\xb_0^{k+1}\|^2
  \notag \\
  &~~~~+\frac{\rho N}{2}\|\xb_0^{k+1}-\xb^\star\|^2,
  %&\geq %-\frac{\rho}{2}\sum_{i=1}^N\|\xb_i^{k+1}-\xb^\star\|_2^2+\frac{\rho}{2}\sum_{i=1}^N\|\xb_i^{k+1}-\xb_0^{k+1}\|^2
\end{align}
and that
\begin{align}\label{eqn: proof of lemma Lc upper bound 7}
  &{\gamma}(\xb_0^{k+1}-\xb_0^k)^T(\xb_0^{k+1}-\xb^\star )=
  \frac{\gamma}{2}\|\xb_0^{k+1}-\xb^\star\|^2\notag \\
  &~~~~-\frac{\gamma}{2}\|\xb_0^{k}-\xb^\star\|^2
  +\frac{\gamma}{2}\|\xb_0^{k+1}-\xb_0^{k}\|^2.
  %&\geq -\frac{\gamma}{2}\|\xb_0^{k}-\xb^\star\|_2^2
  %+\frac{\gamma}{2}\|\xb_0^{k+1}-\xb_0^{k}\|_2^2.
\end{align}
By substituting \eqref{eqn: proof of lemma Lc upper bound 6} and \eqref{eqn: proof of lemma Lc upper bound 7} into
\eqref{eqn: proof of lemma Lc upper bound 512} and recalling $\Lc_\rho$ in \eqref{eqn: Lc}, we obtain
\begin{align}\label{eqn: proof of lemma Lc upper bound 8}
 &\triangle_{k+1} %=\sum_{i=1}^N f_i(\xb_i^{k+1})+ h(\xb_0^{k+1})  +\sum_{i=1}^N(\lambdab^{k +1}_i)^T(\xb_i^{k+1}-\xb_0^{k+1}) +\frac{\rho}{2}\sum_{i=1}^N\|\xb_i^{k+1}-\xb_0^{k+1}\|^2 -  F^\star
 %\notag \\
 \leq \frac{\rho-\sigma^2}{2} \sum_{i=1}^N\|\xb_i^{k+1}-\xb^\star\|^2
 +\frac{\gamma}{2}\|\xb_0^{k}-\xb^\star\|^2
  \notag \\
  &~~~~~-\frac{\gamma}{2}\|\xb_0^{k+1}-\xb_0^{k}\|^2 - \frac{\gamma +\rho N}{2}\|\xb_0^{k+1}-\xb^\star\|^2.
\end{align}

We bound the term $\sum_{i=1}^N\|\xb_i^{k+1}-\xb^\star\|^2$ as %shown in \eqref{eqn: proof of lemma Lc upper bound 9} at the top of the next page, where the last inequality is obtained by assuming $\delta>1$.
%\begin{figure*}
\begin{align}
  &\sum_{i=1}^N\|\xb_i^{k+1}-\xb^\star\|^2 = \sum_{i=1}^N\|\xb_i^{k+1}-\xb_0^{k+1}+\xb_0^{k+1}-\xb^\star\|^2 \notag \\
  &\leq (1+\frac{1}{\delta})N\|\xb_0^{k+1}-\xb^\star\|_2^2+ (1+\delta)\sum_{i=1}^N\|\xb_i^{k+1}-\xb_0^{k+1}\|^2
  ~~~~~~({\sf by~} \eqref{eqn: identity}) \notag 
  %\\
  \end{align}
  \begin{align}
  &\leq (1+\frac{1}{\delta})N\|\xb_0^{k+1}-\xb^\star\|_2^2+ (1+\delta)\bigg[
  \frac{2L^2}{\rho^2}\sum_{i\in \Ac_k}\|\xb_i^{k+1}\!-\xb_i^{k}\|^2 +\frac{2L^2}{\rho^2}\sum_{i\in \Ac_k^c}\|\xb_i^{\widetilde k_i+1}\!-\xb_i^{\widetilde k_i}\|^2  \notag \\
   &~~~~~~+4\sum_{i\in \Ac_k}\|\xb_0^{\bar k_i+1}-\!\xb_0^{k}\|^2+4\sum_{i\in \Ac_k^c}\|\xb_0^{\widehat k_i+1}-\!\xb_0^{k}\|^2+4N\|\xb^{k+1}_0-\xb_0^k\|^2\bigg]~~~~~~({\sf by~} \eqref{eqn: consensus error bound})
   \notag \\
  &  \leq (1+\frac{1}{\delta})N\|\xb_0^{k+1}-\xb^\star\|_2^2
  +  \frac{4\delta L^2}{\rho^2}\sum_{i\in \Ac_k}\|\xb_i^{k+1}\!-\xb_i^{k}\|^2 +\frac{4\delta L^2}{\rho^2}\sum_{i\in \Ac_k^c}\|\xb_i^{\widetilde k_i+1}\!-\xb_i^{\widetilde k_i}\|^2 \notag \\
   &~~~~~~+8\delta\sum_{i\in \Ac_k}\|\xb_0^{\bar k_i+1}-\!\xb_0^{k}\|^2+8\delta\sum_{i\in \Ac_k^c}\|\xb_0^{\widehat k_i+1}-\!\xb_0^{k}\|^2+8\delta N\|\xb^{k+1}_0-\xb_0^k\|^2,
    \label{eqn: proof of lemma Lc upper bound 9}
\end{align}
%\hrulefill
%\end{figure*}
where the last inequality is obtained by assuming $\delta>1$.
%where \eqref{eqn: async cadmm s1 lambda equi2} and \eqref{eqn: async cadmm s1 lambda skc equi} are used to obtain the second inequality and \eqref{lemma: Lc progress eq 8} is used to obtain the last inequality; we have also applied the assumption of $\delta >1$ to the two terms in the RHS of \eqref{eqn: proof of lemma Lc upper bound 8.5}.
Besides, we bound the term $\frac{\gamma}{2}\|\xb_0^{k}-\xb^\star\|^2$ in the RHS of \eqref{eqn: proof of lemma Lc upper bound 8} as
\begin{align}\label{eqn: proof of lemma Lc upper bound 10}
 &\frac{\gamma}{2}\|\xb_0^{k}-\xb^\star\|^2 =
 \frac{\gamma}{2}\|\xb_0^{k}-\xb_0^{k+1}+\xb_0^{k+1}-\xb^\star\|^2 \notag \\
 &\leq \frac{\gamma}{2}(1+\delta)\|\xb_0^{k}-\xb_0^{k+1}\|^2 +  \frac{\gamma}{2}(1+\frac{1}{\delta})\|\xb_0^{k+1}-\xb^\star\|^2.
\end{align}
By substituting \eqref{eqn: proof of lemma Lc upper bound 9} and \eqref{eqn: proof of lemma Lc upper bound 10} into \eqref{eqn: proof of lemma Lc upper bound 8}, one obtains
\begin{align}\label{eqn: proof of lemma Lc upper bound 11}
 &\triangle_{k+1} \leq \bigg(\frac{\rho N + \gamma}{2\delta}
 - \frac{\sigma^2 N}{2}(1+\frac{1}{\delta})\bigg) \|\xb_0^{k+1}-\xb^\star\|^2
 \notag \\
 &~~~~~~~~+\bigg(\frac{\gamma \delta}{2}+{4(\rho-\sigma^2)N \delta }\bigg)\|\xb_0^{k+1}-\xb_0^{k}\|^2 +\frac{2(\rho-\sigma^2)\delta L^2}{\rho^2}\sum_{i\in \Ac_k}\|\xb_i^{k+1}-\xb_i^{k}\|^2\notag \\
 &~~~~~~~~
  + \frac{2(\rho-\sigma^2)\delta L^2}{\rho^2}\sum_{i\in \Ac_k^c}\|\xb_i^{\widetilde k_i+1}-\xb_i^{\widetilde k_i}\|^2
  \notag
   \\
  &~~~~~~~~+4(\rho-\sigma^2)\delta\sum_{i\in \Ac_k}\|\xb_0^{k}-\xb_0^{\bar k_i+1}\|^2
  +4(\rho-\sigma^2)\delta\sum_{i\in \Ac_k^c}\|\xb_0^{k}-\xb_0^{\widehat k_i+1}\|^2.
\end{align}
Let $\delta>1$ be large enough so that
$\frac{\rho N + \gamma}{2\delta} %+\frac{\gamma}{2\delta}
 - \frac{\sigma^2 N}{2}(1+\frac{1}{\delta})\leq 0$
and assume that $\gamma \geq 8({\rho-\sigma^2})N$. Then, one obtains \eqref{eqn: lemma LC upper bound 0} from \eqref{eqn: proof of lemma Lc upper bound 11}.

To show \eqref{eqn: lemma LC upper bound 0 gamma0}, let $\gamma=0$ in \eqref{eqn: proof of lemma Lc upper bound 11} and assume that $\delta>1$ be large enough so that
$\frac{\rho}{\delta} %+\frac{\gamma}{2\delta}
 - {\sigma^2}(1+\frac{1}{\delta})\leq 0$.
 \hfill $\blacksquare$

\section{Proof of Lemma \ref{lemma: osc}}\label{appx: proof of lemma: osc}
%{\bf Proof of Lemma \ref{lemma: osc}:}
Since $\Xc^\star$ is a linear set, according to the Hoffman bound \cite{Hoffman52}, for some constant $c>0$, %{\red [[I don't think Hoffman's bound can be applied. we have discussed this before?]]}
\begin{align}\label{eqn: hoffman bound}
&{\rm dist}^2(\Xc^\star,\hat \xb)=\sum_{i=1}^N\|\Pc^\star(\hat \xb)-\xb_i\|^2 + \|\Pc^\star(\hat \xb)-\xb_0\|^2 \notag\\
&~~~~~~~\leq c \sum_{i=1}^N\|\Ab_i\xb_i-\yb_i^\star\|^2 + c \sum_{i=1}^N \|\xb_i-\xb_0\|^2.
\end{align}
In addition, it follows from the strong convexity of $g_i$'s that
\begin{align}\label{eqn: OSC 4}
  &\sum_{i=1}^N f_i(\Pc^\star(\hat \xb)) =\sum_{i=1}^N g_i(\Ab_i\Pc^\star(\hat \xb)) \notag
  \\
  %\end{align}
  %\begin{align}
  &\geq  \sum_{i=1}^N g_i(\Ab_i\xb_i)+ \sum_{i=1}^N (\nabla g_i(\Ab_i\xb_i))^T\Ab_i(\Pc^\star(\hat \xb)-\xb_i)+  \sum_{i=1}^N\frac{\sigma^2}{2}\|\Ab_i\Pc^\star(\hat \xb)-\Ab_i\xb_i\|^2
  \notag
  \\
  %\end{align}
  %\begin{align}
  &=\sum_{i=1}^N f_i(\xb_i)+ \sum_{i=1}^N (\nabla f_i(\xb_i))^T(\Pc^\star(\hat \xb)-\xb_i) +  \sum_{i=1}^N\frac{\sigma^2}{2}\|\yb_i^\star-\Ab_i\xb_i\|^2. %\notag \\
  %&\geq \sum_{i=1}^N f_i(\xb_i)+ \sum_{i=1}^N (\nabla f_i(\xb_i))^T(\Pc^\star(\hat \xb)-\xb_i) +  \sum_{i=1}^N\frac{\sigma^2}{2c}\|\Pc^\star(\hat %\xb)-\xb_i\|^2 \notag \\
  %&~~~~~+\frac{\sigma^2}{2c}\|\Pc^\star(\hat \xb)-\xb_0\|^2 - \frac{\sigma^2}{2}\sum_{i=1}^N\|\xb_0-\xb_i\|^2,
\end{align}
By substituting \eqref{eqn: hoffman bound} into \eqref{eqn: OSC 4}, one obtains \eqref{eqn: OSC 3}.
\hfill $\blacksquare$

\section{Proof of Lemma \ref{lemma: LC upper bound 2}}\label{appx: proof of lemma: LC upper bound 2}
%{\bf Proof of Lemma \ref{lemma: LC upper bound 2} :}
By applying \eqref{eqn: OSC 3} (with $\xb_i=\xb_i^{k+1}$ $\forall i=0,1,\ldots,N$) to \eqref{eqn: proof of lemma Lc upper bound 510}, and following the same steps as in \eqref{eqn: proof of lemma Lc upper bound 512}-\eqref{eqn: proof of lemma Lc upper bound 8}, we have
\begin{align}\label{eqn: proof of lemma Lc upper bound 2 8}
 &\triangle_{k+1} %=\sum_{i=1}^N f_i(\xb_i^{k+1})+ h(\xb_0^{k+1})  +\sum_{i=1}^N(\lambdab^{k +1}_i)^T(\xb_i^{k+1}-\xb_0^{k+1}) +\frac{\rho}{2}\sum_{i=1}^N\|\xb_i^{k+1}-\xb_0^{k+1}\|^2 -  F^\star
 %\notag \\
 \leq \frac{\rho-\sigma^2/c}{2} \sum_{i=1}^N\|\xb_i^{k+1}-\Pc^\star(\hat \xb)\|^2
 +\frac{\gamma}{2}\|\xb_0^{k}-\Pc^\star(\hat \xb)\|^2
  \notag \\
  &~~~~-\frac{\gamma}{2}\|\xb_0^{k+1}-\xb_0^{k}\|^2 - \frac{\gamma +\sigma^2/c+\rho N}{2}\|\xb_0^{k+1}-\Pc^\star(\hat \xb)\|^2 \notag \\
  &~~~~+\frac{\sigma^2}{2}\sum_{i=1}^N\|\xb_i^{k+1}-\xb_0^{k+1}\|^2.
\end{align}

Recall \eqref{eqn: proof of lemma Lc upper bound 9}, \eqref{eqn: proof of lemma Lc upper bound 10} (with $\xb^\star$ replaced by $\Pc^\star(\hat \xb)$) and \eqref{eqn: consensus error bound} in Lemma \ref{lemma: consensus error} and apply them to
\eqref{eqn: proof of lemma Lc upper bound 2 8}. One obtains
\begin{align}\label{eqn: proof of lemma Lc upper bound 2 11}
 &\triangle_{k+1} \leq \bigg(\frac{\rho N + \gamma}{2\delta}
 - \frac{N\sigma^2/c }{2}(1+\frac{1}{\delta})\bigg) \|\xb_0^{k+1}-\Pc^\star(\hat \xb)\|^2
 \notag \\
 &~~~~~+\bigg(\frac{\gamma \delta}{2}+{4(\rho-\sigma^2/c)N \delta +2\sigma^2 N}\bigg)\|\xb_0^{k+1}-\xb_0^{k}\|^2 \notag \\
 &~~~~~+\frac{(2(\rho-\sigma^2/c)\delta +\sigma^2) L^2}{\rho^2}\sum_{i\in \Ac_k}\|\xb_i^{k+1}-\xb_i^{k}\|^2
 \notag
 %\\
 \end{align}
 \begin{align}
 &~~~~~
  + \frac{(2(\rho-\sigma^2/c)\delta +\sigma^2) L^2}{\rho^2}\sum_{i\in \Ac_k^c}\|\xb_i^{\widetilde k_i+1}-\xb_i^{\widetilde k_i}\|^2
  \notag
   \\
  &~~~~~+(4(\rho-\sigma^2/c)\delta+2\sigma^2)\sum_{i\in \Ac_k}\|\xb_0^{k}-\xb_0^{\bar k_i+1}\|^2
  \notag\\
  &~~~~~+(4(\rho-\sigma^2/c)\delta+2\sigma^2)\sum_{i\in \Ac_k^c}\|\xb_0^{k}-\xb_0^{\widehat k_i+1}\|^2.
\end{align}
Let $\delta>1$ be large enough so that
$\frac{\rho N + \gamma}{2\delta} %+\frac{\gamma}{2\delta}
 - \frac{N\sigma^2/c }{2}(1+\frac{1}{\delta})\leq 0$
In addition, since $\gamma \geq 8N{(\rho-\sigma^2/c)+4N\sigma^2}$ implies $\gamma \geq 8N{(\rho-\sigma^2/c)+4N\sigma^2/\delta}$, \eqref{eqn: proof of lemma Lc upper bound 2 11} infers \eqref{eqn: lemma LC upper bound 0}.

To obtain \eqref{eqn: lemma LC upper bound 0 gamma02}, let $\gamma=0$ in \eqref{eqn: proof of lemma Lc upper bound 2 11} and assume that $\delta>1$ be large enough so that
$\frac{\rho}{\delta} %+\frac{\gamma}{2\delta}
 - \frac{\sigma^2}{c}(1+\frac{1}{\delta})\leq 0$.
\hfill $\blacksquare$

\vspace{-0.0cm}
\footnotesize
\bibliography{distributed_opt,smart_grid,ref}
\end{document}